\documentclass{aa}
\usepackage{graphicx} % Required for inserting images
\usepackage{txfonts}
\usepackage{epsfig}
\usepackage{amsbsy}
\usepackage{amssymb}
\usepackage{amsmath}
\usepackage{latexsym}
\usepackage{natbib}
\usepackage{verbatim}
\usepackage{bm}
\usepackage{tabularx}
\usepackage{soul}
\usepackage{xcolor}
\usepackage{lastpage}

\usepackage[
  breaklinks = true,
  colorlinks = true,
  urlcolor   = blue,
  citecolor  = blue,
  linkcolor  = blue,
]{hyperref}
\definecolor{darkgreen}{rgb}{0,0.60,0}

\definecolor{cyan}{rgb}{0,0.8,0.8}

\definecolor{olive}{rgb}{0.5,0.5,0}

\usepackage[nonumberlist,nosuper]{glossaries}
\setacronymstyle{long-short}
\newacronym{3d}{3D}{three-dimensional}
\newacronym{cbp}{CBP}{coronal bright points}
\newacronym{clv}{CLV}{center-to-limb variation}
\newacronym{los}{LoS}{line of sight}
\newacronym{pos}{PoS}{plane of sky}
\newacronym{rmhd}{r-MHD}{radiative magnetohydrodynamic} 
\newacronym{rt}{RT}{radiative transfer}
\newacronym{see}{SEE}{statistical equilibrium equations}
\newacronym{snr}{SNR}{signal-to-noise ratio}
\newacronym{wfa}{WFA}{weak field approximation}
\begin{document}

\title{Spectropolarimetric synthesis of forbidden lines \\ in MHD models of coronal bright points}

\titlerunning{Polarization of forbidden lines in CBPs}

\authorrunning{Alsina Ballester, N\'obrega-Siverio, Moreno-Insertis  \& Supriya} % Trujillo Bueno}

\author{
Alsina Ballester, E.\inst{1,2}
\and
N\'{o}brega-Siverio, D.\inst{1,2,3,4}
\and
Moreno-Insertis, F. \inst{1,2}
\and
Supriya, H.~D. \inst{1,2}
%\and
%Trujillo Bueno, J. \inst{1, 2, 5}
}

\institute{
Instituto de Astrof\'isica de Canarias, E-38205 La Laguna, Tenerife, Spain
	\and
	Departamento de Astrof\'isica, Universidad de La Laguna, E-38206 La Laguna, Tenerife, Spain
    \and
    Rosseland Centre for Solar Physics, University of Oslo, PO Box 1029 Blindern, 0315 Oslo, Norway
    \and
    Institute of Theoretical Astrophysics, University of Oslo, PO Box 1029 Blindern, 0315 Oslo, Norway
%    \and
%    Consejo Superior de Investigaciones Cient\'{i}ficas, Spain
}

\abstract
{The inference of the magnetic field vector from spectropolarimetric observations is 
crucial for understanding the physical processes governing the solar corona.} 
{We investigate which information on the magnetic fields of coronal bright points (CBP) can be gained  from the
intensity and polarization of the Fe~{\sc{xiii}} $10747$~\AA , Fe~{\sc{xiv}} $5303$~\AA , Si~{\sc{x}} $14301$~\AA , 
and Si~{\sc{ix}} $39343$~\AA\ forbidden lines.} 
{We apply the P-CORONA synthesis code to a CBP model in the very low corona, obtained with the \textit{Bifrost} code, and to a larger global model to study the impact of the outer coronal material along the line of sight (LoS).} 
{The enhanced density within the CBP produces an intensity brightening, but suppresses the linear polarization. The circular polarization from such regions often approaches $0.1\%$ of the intensity. The relative contribution from the coronal material along the LoS depends strongly on its temperature, and is weaker for lines with a peak response at higher temperatures ({Fe~\sc{xiii} $10747$~\AA } at $1.7$~MK and {Fe~{\sc{xiv}} $5303$~\AA } at $2$~MK). The weak field approximation (WFA) provides information on the longitudinal magnetic fields in the strongest-emitting spatial intervals along the LoS, and we find it to be more reliable in the regions of the CBP where the field does not change sign. This tends to coincide with the regions where there is a strong correlation between the circular polarization and the wavelength derivative of the intensity. Considering roughly 30 minutes of time evolution, the CBP signals are somewhat attenuated but are still clearly identifiable, and the area where the WFA can be suitably applied remains substantial.}
{The circular polarization of the {Fe~{\sc{xiv}} $5303$~\AA\ }and especially {Fe~{\sc{xiii}} $10747$~\AA\ }lines are valuable diagnostics for the magnetic fields in the higher-temperature regions of the CBP, which could be exploited with future coronagraphs with similar capabilities to Cryo-NIRSP/DKIST, but designed to observe below $1.05~R_\odot$.} 
\keywords{Sun: corona, Magnetohydrodynamics (MHD), Radiative transfer, Polarization}

\maketitle

\section{Introduction}
\label{sec::Intro}
Despite the importance of magnetism in the physics of the solar corona \citep[e.g.,][]{BPriest, Raouafi+16, Green+18, BJudgeIonson24}, 
measurements of its magnetic fields remain very scarce. 
Although observations of the radiation intensity can provide 
valuable information, 
often from extreme ultraviolet (EUV) lines and relying on extrapolations 
\citep[e.g.,][]{Warren+18, Jarolim+23, Madjarska+24}, 
quantitative data are much more readily available through the polarization of spectral 
lines \citep[e.g.,][]{TrujilloBuenodelPinoAleman22}. 
Remarkably, the spectral lines that arise from transitions that are not allowed by the electric-dipole selection 
rules (i.e., forbidden lines) are of particular interest for probing the corona. 
The most prominent of these lines 
are those produced by magnetic-dipole-type 
(M$1$) transitions between fine-structure levels of the ground term 
of the corresponding chemical species \citep{Judge98}. 
As such, these lines have relatively long wavelengths, typically falling in the visible or infrared range of the spectrum. 

The polarization patterns of such lines are sensitive to the magnetic field through two main physical mechanisms: 
the Hanle and Zeeman effects. 
For forbidden lines emitted in the rarefied corona, both collisional and radiative processes can contribute significantly 
to populating the upper level of the considered transition \citep{SahalBrechot74, Judge98, CasiniJudge99}. 
Collisions excite the atoms into high-energy states, from which they cascade down to the 
transition's upper level. 
The absorption of the radiation from the solar photosphere also contributes to populate the upper level and, 
because such radiation field is anisotropic, 
population imbalances and quantum coherence arise between the magnetic sublevels (i.e., atomic polarization is produced). 
In particular, if the radiation induces population imbalances between sublevels with different quantum number $|M|$ 
(i.e., atomic alignment), the emitted radiation is linearly polarized, in a process known as scattering polarization. 
Furthermore, the magnetic field tends to modify the atomic polarization 
by reducing the quantum interference between magnetic sublevels through the Hanle effect. 
Thus, the magnetic field leaves its signatures 
on the scattering polarization, usually by depolarizing it and rotating the plane of linear polarization  
\citep[for an in-depth discussion on scattering polarization and
the Hanle effect, see][]{TrujilloBueno+01}. 
The Hanle effect operates when the Larmor frequency (which is proportional to the splitting 
between magnetic sublevels) is comparable 
to the inverse lifetime of the level. 
For the considered forbidden lines, this 
occurs for fields of the order of $10^{-6}$~G or weaker. 
In the solar corona, where magnetic fields on the order of gauss are commonplace, such lines are in the Hanle saturation regime, 
and thus their scattering polarization encodes information only on the orientation of the magnetic field, but not its strength. 
Magnetic fields can also produce polarization signals through the Zeeman effect, by causing an energy splitting between 
the magnetic sublevels of the atomic levels involved in the transition. 
This occurs even in the absence of atomic level polarization, so both collisional processes 
and radiative absorption contribute to the Zeeman polarization patterns. 
The amplitude of the signals produced by the Zeeman effect scale with the ratio of
the magnetic splitting to the width of the line, $\mathcal{R}$. The ratio increases linearly with the line's 
wavelength and decreases with the square root of the temperature. 
For the typical temperatures and magnetic fields of the corona, this quantity is generally quite small, 
even for infrared lines with long wavelengths (taking values of $\sim\!10^{-3} - 10^{-4}$). 
As such, linear polarization Zeeman signals (which scale with $\mathcal{R}^2$) 
should be vanishingly small, but non-negligible  circular polarization signals
{(which scale linearly with $\mathcal{R}$)} could be expected. Indeed, the detection of the latter signals was 
recently reported in \cite{Schad+24}. 

In the present work, we study the potential of the polarization of forbidden spectral lines for diagnostics of 
magnetic fields in fundamental structures of the lower solar corona{, namely \gls*{cbp}}. The four specific 
forbidden lines that we study are the green Fe~{\sc{xiv}} line at $5303$~\AA\ (hereafter Fe$5303$), the Fe~{\sc{xiii}} 
line at $10747$~\AA\ (Fe$10747$), the Si~{\sc{x}} line at $14301$~\AA\ (Si$14301$), and the Si~{\sc{ix}} line at $39343$~\AA\ (Si$39343$). 
The interest in these lines is highlighted by the multitude of observations that 
have been carried out in the past, especially in linear polarization 
\citep[e.g.,][]{EddyMcKimMalville67, Mickey73, Arnaud82, Tomczyk+07, Dima+19}. 
In addition, in a select few investigations, measurements of the coronal magnetic field were
obtained from observations in circular polarization \citep[see][]{Lin+04, Schad+24}.
Recent instruments for coronal spectropolarimetry have been designed specifically for the 
wavelengths corresponding to some or all of these lines, including SOLARC \citep{Kuhn+03}, 
UCoMP (\citealt{Tomczyk+21}; see also \citealt{Tomczyk+08}), VELC/Aditya-L1 \citep{Singh+19}, or Cryo-NIRSP/DKIST \citep{Rimmele+20,Fehlmann+23}. 

In recent years, a number of theoretical investigations have been carried out on these lines, based on spectral synthesis using 
suitable atmospheric models. These have been enabled by the development of numerical codes that can account for \gls*{3d} atmospheric 
models and for the relevant physics (including collisional processes, scattering polarization, and the Hanle and Zeeman effects), while 
the medium can be treated as optically thin. For instance, the FORWARD toolset \citep{Gibson+16} has been used to investigate a variety of 
permitted coronal lines \citep[e.g.,][]{Khan+23}, as well as forbidden lines, including Fe$10747$ \citep[e.g.,][]{Karna+19}. \cite{Li+17} 
used the density matrix formalism for multi-level atoms under the 
assumption of a flat-spectrum incident radiation field \citep{CasiniJudge99, CasiniJudge00, BLandiLandolfi04}. 
In the same paper, the intensity and polarization signals of the 
Fe$10747$ line were synthesized considering potential-field source-surface coronal magnetic field models. 
The PyCELP code \citep[see][]{SchadDima20,SchadDima21}, 
based on the same formalism, was recently applied to models obtained with the radiative magnetohydrodynamic (r-MHD) MURaM code, representative of active 
regions \citep{Rempel17}. Such investigations considered six lines of interest for Cryo-NIRSP/DKIST, including the four lines studied in this work
in addition to {the} Fe~{\sc{xi}} line at $7892$~\AA\ and {the} Fe~{\sc{xiii}} line at $10798$~\AA . 
The recently developed and publicly available P-CORONA code can 
suitably treat both forbidden and permitted spectral lines \citep{Supriya+21, Supriya+25}. Moreover, unlike other previously mentioned codes, P-CORONA does not assume the weak field limit when accounting for the impact of the Zeeman effect. A detailed description of this code can be found in \cite{Supriya+25}. That 
article also presents brief illustrative applications of the code to the same six lines investigated in \cite{SchadDima20,SchadDima21}, but considering atmospheric models 
representative of the full extended corona as well as various models obtained with the MURaM code representing different low-corona conditions.

In the present work{, we} use P-CORONA to examine in detail the spectropolarimetric signals of the 
Fe$5303$, Fe$10747$, Si$14301$, and Si$39343$ lines 
for the specific case of CBPs. These low-coronal structures are ubiquitous in the Sun, 
regardless of the level of activity \citep{Madjarska19}, and are 
{possibly important contributors to coronal heating} \citep{Priest+94}. 
\gls*{cbp}s, composed of small 
loops that confine plasma heated to several million degrees, can be considered small-scale analogues of 
active regions \citep{Gao+22}. To investigate the polarization signals of forbidden lines produced in a \gls*{cbp}, 
we compute them numerically with P-CORONA, considering the models resulting from \gls*{rmhd} 
simulations with the \textit{Bifrost} code \citep{Gudisken11} carried out by \cite{NobregaSiverio+23}. 
Present-day coronagraphs such as Cryo-NIRSP/DKIST are not designed for 
observations less than $0.05\,R_\odot$ from the base of the corona, 
where most \gls*{cbp}s are expected to be found, since their {projected} size ranges from about $5$ to $40$~Mm{ and their
heights range between $5$ and $10$~Mm \citep{Madjarska19}}. 
Indeed, {the main} goal of 
{this} investigation is to evaluate the suitability of 
techniques based on the polarization of forbidden lines for studying \gls*{cbp} magnetic fields, 
thereby providing insight for future instrumental developments and observational studies. 

In Section~\ref{sec::Formulation}, we present the methodology that we employed, the atmospheric models that we considered, 
and the main assumptions that we made. In Section~\ref{sec::Stokes}, we show the synthetic wavelength-integrated 
intensity and polarization images corresponding to emission from a \gls*{cbp}.
In Section~\ref{sec::LoSMat}, we analyze how the same images are impacted by coronal material 
{lying} along the \gls*{los} direction. In Section~\ref{sec::WFA}, we investigate the 
suitability of inferring the magnetic fields in the \gls*{cbp} from the 
circular polarization signals {of the emitted radiation}. 
In Section~\ref{sec::Timint}, we consider the intensity and polarization signals produced 
when accounting for the time evolution of \gls*{cbp}. The overall conclusions are presented in Section~\ref{sec::Conclusions}. 

\section{Formulation}
\label{sec::Formulation}

The four forbidden lines that we analyze in this work arise from M$1$ transitions between 
different $J$ levels of the ground term of their corresponding chemical species. 
The atomic parameters for the lines of interest are shown in Table~\ref{Tab::Table1}, adapted from 
Table 1 of \cite{SchadDima20} and reproduced here for the reader's convenience. 
Each line probes the coronal material at a different temperature, having {their peak response} 
at temperatures between $1.1$~and~$2$~MK.\footnote{Such temperatures are largely determined by the ionization 
fraction of the relevant chemical species, while the  population of the line's upper level relative 
to that of the ion varies much more slowly with temperature.} 
For all four lines, their upper levels can carry atomic alignment (thus producing scattering 
polarization) and have very long lifetimes, with the corresponding Einstein coefficients 
for spontaneous emission being on the order of $10^{-1}$ to $10$ s$^{-1}$. 
Thus, they are subject to the Hanle effect for magnetic fields on the order of $10^{-6}$~G or weaker, 
and Hanle saturation can be safely assumed throughout the entire corona.
As noted in Section~\ref{sec::Intro}, circular polarization can also be produced in these lines through the Zeeman effect.

\begin{table}[h!]
\centering
\caption{Spectral line properties, adapted from Table~1 of \cite{SchadDima20}.}
\begin{tabular}{cccc}
\hline
Line [\AA] (air) & Transition & $A_{u \ell}$ & $\log(T_\mathrm{max}/[K])$ \\
\hline\hline
Fe~{\sc{xiv}} $5303$ & $3s^2\,3p\,{}^2\!P_{3/2\to1/2}$  & $55.2$ & $6.3$ \\
Fe~{\sc{xiii}} $10747$ & $3s^2\,3p^2\,{}^3\!P_{1\to0}$ & $14.0$ & $6.25$ \\
Si~{\sc{x}} $14301$ & $2s^2\,2p\,{}^2\!P_{3/2\to1/2}$ & $3.08$ & $6.15$ \\
Si~{\sc{ix}} $39343$ & $2s^2\,2p^2\,{}^3\!P_{3/2\to1/2}$ & $0.30$ & $6.05$ \\
\hline
\end{tabular}
\label{Tab::Table1}
\end{table}

\subsection{The P-CORONA synthesis code}
\label{sec::FormulationSub_PCORONA}
In this work, the intensity and polarization profiles of the above-mentioned
spectral lines were primarily computed with the P-CORONA code \citep{Supriya+21,Supriya+25}.  
This {spectral synthesis} code is suitable for any %three-dimensional
\gls*{3d} model of the solar corona, 
and it assumes that the medium is optically thin and is illuminated by the underlying 
unpolarized photospheric radiation field. 
At each spatial grid point, the wavelength-dependent polarized emissivity is computed by solving the \gls*{see}. 
Such equations incorporate both the collisional and radiative coupling between the atomic levels of the system. 
They account for the atomic polarization in each level, which is ultimately produced by 
the anisotropic photospheric 
continuum radiation. In this work, we used the photospheric radiation field intensities 
provided in the public 
version of P-CORONA. We also accounted for the \gls*{clv} of the radiation field, taking the limb-darkening 
coefficients given in \cite{BAllen02}. 

For simplicity, the assumption of complete frequency redistribution is made when solving the \gls*{see}, according to 
which the frequencies of incident and emitted radiation are treated as uncorrelated in scattering processes. 
This is a reasonable assumption in the considered scenario, in which the illuminating radiation field 
can be treated as spectrally flat in the vicinity of the spectral lines of interest. 
The impact of the magnetic field on the emission vector is taken into account through both the Hanle and Zeeman effects. 
In P-CORONA, the velocities can be taken into account both in the \gls*{see} due to the symmetry-breaking effects of 
their non-radial component and through the Doppler shift they induce in the emitted radiation. 

The intensity and polarization profiles of the emergent radiation at each point on the \gls*{pos} are computed by spatially integrating 
the polarized emissivities along the \gls*{los} direction, using the trapezoidal rule. 
This integration is performed separately for each discrete vacuum wavelength point ($\lambda_i$), 
yielding Stokes profiles $I(\lambda_i)$, $Q(\lambda_i)$, $U(\lambda_i)$, and $V(\lambda_i)$. 
We took a wavelength grid with $301$ points for each considered line, spanning a wavelength range of $10$~\AA\ for the 
Fe lines and of $22$~\AA\ for the Si lines. 

In the following sections, we display the wavelength-integrated profiles, 
{also obtained} using trapezoidal integration. %
We write the wavelength-integrated intensity and linear polarization signals as $\overline{I\,}$, $\overline{Q\,}$, and 
$\overline{U\,}$. Because of the typical antisymmetric shape of the Stokes $V(\lambda)$ profiles, 
we show the wavelength-integration of its absolute value, $\overline{|V|}$, to avoid cancellations within the integration range. 
Rather than showing the linear polarization in terms of $\overline{Q\,}$ and $\overline{U\,}$, 
in the following sections we represent them in terms of the linear polarization fraction $P_L$, 
\begin{equation}
    P_L = \frac{\sqrt{\overline{Q\,}^{\,2} + \overline{U\,}^{\,2}} }{\overline{I\,}} \, ,
\end{equation}
and the linear polarization angle $\alpha$, defined as the angle between the direction of maximum linear polarization and the vertical ($Z$) axis, 
\begin{align}
\label{Eq::PolAng}
   \alpha = \left\{
        \begin{array}{cc}
         \quad \frac{1}{2} \, \cos^{-1} \biggl(\overline{Q} \, \biggl/ \sqrt{\overline{Q}^{\,2} + \overline{U}^{\,2}}  \biggr) & \quad \overline{U} > 0 \; , \\ 
           \, -\frac{1}{2} \, \cos^{-1} \biggl(\overline{Q} \, \biggl/ \sqrt{\overline{Q}^{\,2} + \overline{U}^{\,2}}  \biggr) & \quad \overline{U} \leq 0 \; .
        \end{array}   
   \right.
\end{align}

\subsection{Atmospheric models}
 \label{sec::FormulationSub_Bifrost}
 
 \subsubsection{The CBP model}
 \label{sec::FormulSubAtmosSSubCBP}
In order to study the observable polarization and intensity signals resulting 
from \gls*{cbp} structures of the low corona, 
we carry out the synthesis from the numerical model by \cite{NobregaSiverio+23} obtained with the \gls*{rmhd} 
\textit{Bifrost} code. This \gls*{3d} model represents a \gls*{cbp} embedded within a coronal hole. 
The simulation box for this model, with a height of $34.2$~Mm and a horizontal extent of 
$32^2$~Mm$^2$, is composed of $512^3$ cells. 
Except where otherwise noted, we take the model at a typical instant of the evolution of the \gls*{cbp}, 
namely at $201.7$ minutes from the start of the {simulation}.

In the calculations presented in the following sections, the coronal elemental abundances were taken 
from \cite{Schmelz+12} 
which are taken as default in the atomic models provided in the public version of P-CORONA. For Fe and Si, these 
abundances are $7.85$ and $7.86$, respectively.\footnote{Such abundances are lower than the 
ones obtained when accounting for first ionization potential (FIP) effects \citep[e.g.,][]{Laming15}. 
Accounting for such increased abundances should not be expected to substantially affect the polarization fraction of the 
considered lines, 
{but could enhance their intensity}. 
Thus, the intensities presented in the following sections can be taken as a lower limit, implying that 
the observed signals should be easier to detect.} 
The proton densities are taken to be equal to the electron densities of the atmospheric model times a proportionality factor which depends on the assumptions for the chemical composition made in the model.  
Assuming a solar coronal composition and full ionization (as is the case for the \textit{Bifrost} models), this factor is $0.85$.

\subsubsection{Enveloping coronal material}
\label{sec::FormulSubAtmosSSubPS}
In the present work, we also analyze how the coronal material lying on the \gls*{los}, beyond the relatively small extent 
of $32$~Mm (or $0.046\,R_\odot$) of the \gls*{cbp} model, impacts the observable signal. 
For this purpose, we consider an atmospheric model included in the public version of P-CORONA{: a global MHD model generated} 
using solar data corresponding to Carrington rotation $2118$ and longitude $66^\circ$ from the 
``MHDweb Modeling of Solar Corona \& Heliosphere'' project\footnote{\url{https://www.predsci.com/mhdweb/home.php}} of Predictive Science, Inc. 
The full model (hereafter PSI) has $151^3$ spatial cells and extends from $-3\,R_\odot$ to $3\,R_\odot$ in the 
\gls*{los} dimension 
and from $-1.5\,R_\odot$ to $1.5\,R_\odot$ in each of the dimensions of the \gls*{pos}. 
The PSI models considered here assume a fully ionized gas of pure hydrogen \citep[e.g.,][]{Riley+11}. In this case, 
the ratio of proton to electron density is $1$. 

Under the assumption that the coronal material {is} optically thin, summing the profiles obtained 
through the P-CORONA calculations both from the \gls*{cbp} model and from the PSI model 
represents a good approximation to the radiation that is emitted along the \gls*{los} crossing the \gls*{cbp} 
and the surrounding material. 

\begin{figure}[!h]
 \centering
 \includegraphics[width=0.36\textwidth]{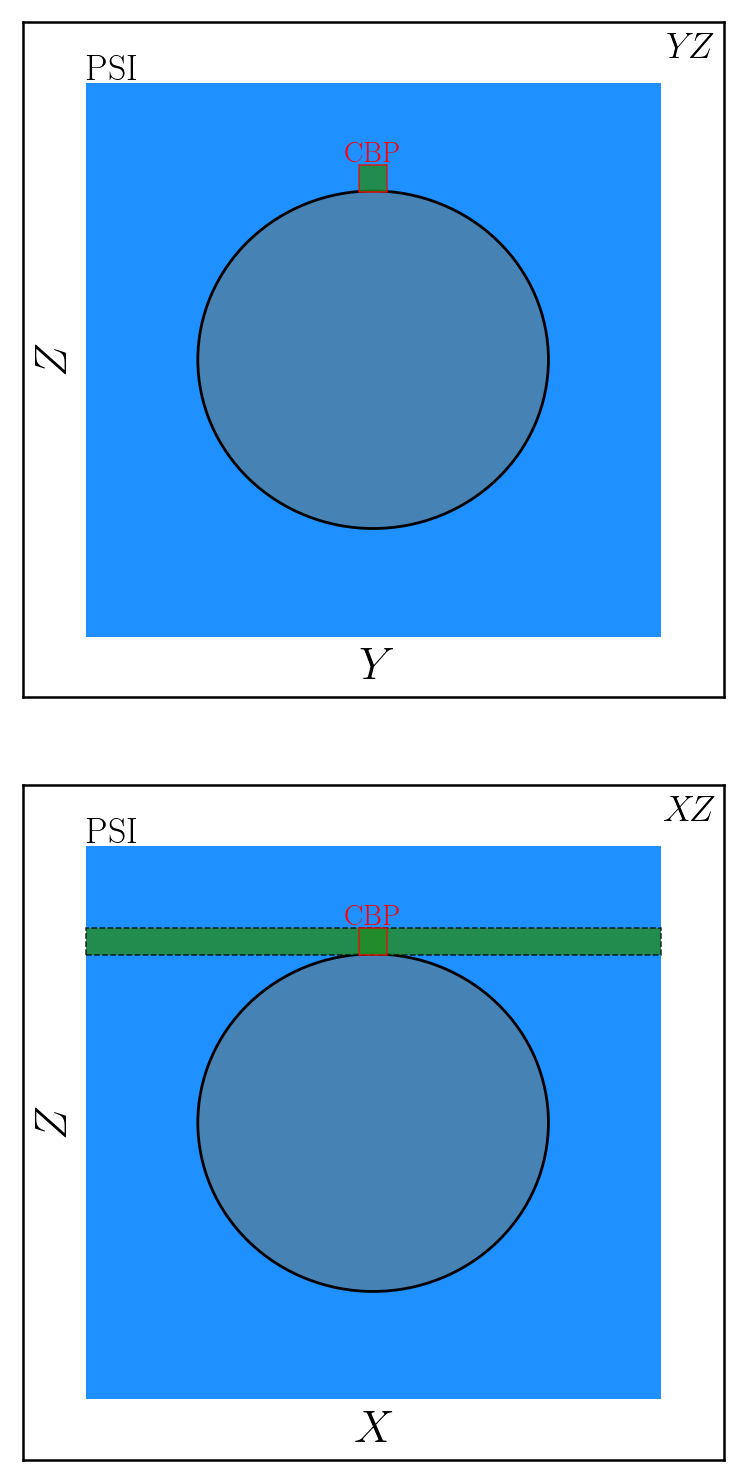}
 \caption{{Diagram illustrating the shape of the column taken from the original PSI model, considered in this work. The blue shaded area corresponds to the original 
 PSI model. The green shaded area corresponds to the selected column. The red square shows the boundary of the CBP model. The black circle 
 shows the boundary between the corona and the inner solar atmosphere. \textit{Upper panel}: $YZ$ plane at $X = 0$, with the $X$-direction 
 corresponding to the \gls*{los}. \textit{Lower panel}: $XZ$ plane for $Y = 0$. Figure not to scale.}}
 \label{fig::Diagram}
\end{figure} 
In order to sum the {the profiles resulting from both models, 
such models must extend over the
same area on the \gls*{pos} (i.e., the $YZ$ plane). Thus, we take columns from the PSI model
that extend the full \gls*{los} range (i.e., along the $X$ direction) but we cut in the 
$Y$ and $Z$ dimensions so that they extend over $32$~Mm (or $\sim\!0.046\,R_\odot$) and $34.2$~Mm (or $\sim\!0.049\,R_\odot$), 
respectively. A diagram illustrating the shape of such columns is shown in Figure~\ref{fig::Diagram}. 
In order for
the cells in these columns to have the same \gls*{pos} dimensions as those of the \gls*{cbp}
model, so that the profiles from both models can be suitably summed, we further refined the spatial
grid of the columns of the PSI model so that they have $512^2$ cells on the \gls*{pos}.}
We consider the contribution to the signals from two different columns of the PSI model. 
The first column that we took touches the base of the corona at the solar North pole 
(with $0.995 \,R_\odot \lesssim Z \lesssim 1.05\,R_\odot$, hereafter model PSI1). 
The other column touches the solar South pole and is inverted along the $Z$-axis 
(with $-1.05\,R_\odot \lesssim Z \lesssim -0.995\,R_\odot $, hereafter model %B)
PSI2). 
{With these choices, the $Z$ coordinates very nearly coincide with the height 
just above the photosphere}. 
Because of the low resolution of the original full PSI model compared to the \gls*{cbp} model, 
after interpolation PSI1 and PSI2 present very little variation along the $Y$ axis and a smooth variation along $Z$. 

{We note that, through this approach, the contribution from the regions along the LoS between 
$X \simeq -0.023 R_\odot$ to $X \simeq 0.023 R_\odot$ is counted twice, once for each model.} 
{Regardless, 
this approach avoids the discontinuities in the magnetic field that would be 
introduced if the two models were merged while removing the part of the PSI section that coincides with the 
\gls*{cbp} model in the $X$ direction. Moreover, we verified that the contribution to the overall intensity
from the overlapping region of either of the PSI models is not appreciable at the plot level, for any of the considered lines.} 
The intensity and polarization signals obtained by accounting for the joint contributions from
the \gls*{cbp} and PSI models are shown and discussed in Section~\ref{sec::LoSMat},
after presenting the results of the \gls*{cbp} model {alone} in Section~\ref{sec::Stokes}.  

\begin{figure*}[!t]
 \centering
 \includegraphics[width=0.99\textwidth]{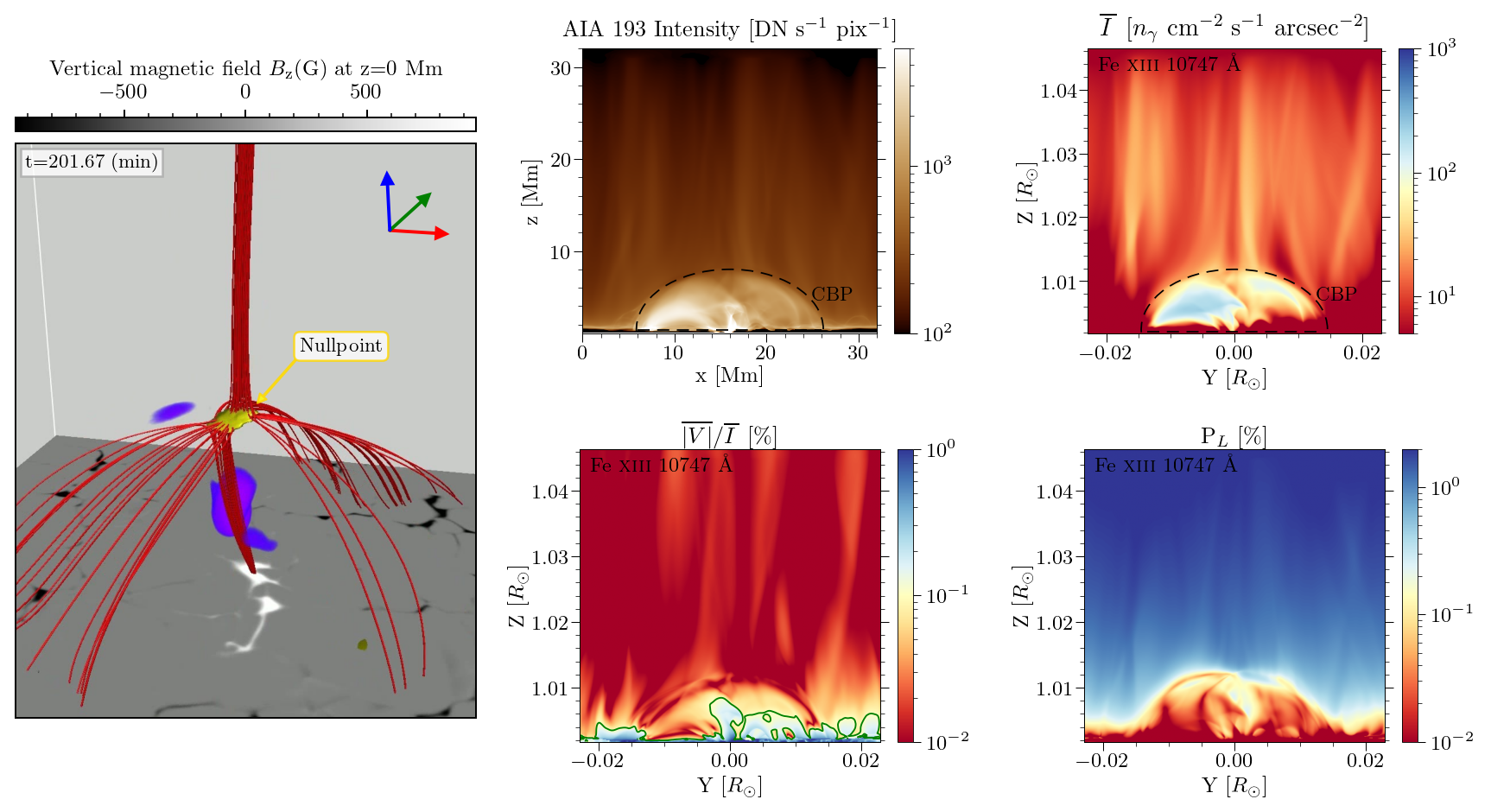}
 \caption{\gls*{cbp} model corresponding to $201.7$ minutes from the start of the \textit{Bifrost} simulation
 originally presented in \cite{NobregaSiverio+23}, and associated spectral synthesis calculations. 
 \textit{Left panel}: Magnetic topology of the 
 \gls*{cbp} model (red lines), superimposed on a horizontal map of $B_\mathrm{z}$ at $\mathrm{z} = 0$. 
 The red-green-blue coordinate system of the \textit{Bifrost} model indicates the 
 $\mathrm{x}, \mathrm{y}, \mathrm{z}$ axes. 
 \textit{Upper central panel}: Synthetic intensity computed with CHIANTI \citep{Dere+23}, mimicking 
a limb observation by SDO/AIA $193$~\AA . 
The rest of the panels correspond to synthetic quantities of Fe$10747$ line, obtained with P-CORONA. 
\textit{Upper right panel}: Wavelength-integrated intensity $\overline{I\,}$. 
\textit{Lower central panel}: Wavelength-integrated absolute-value circular polarization, normalized to the integrated intensity, $\overline{|V|}/\overline{I\,}$. 
\textit{Lower right panel}: Wavelength-integrated linear polarization fraction $P_L$. 
The dashed black curves in the upper central and upper right panels indicate
the region corresponding to the \gls*{cbp}. 
The green contour in the lower central panel delineates where the $\overline{|V|}/\overline{I\,}$ 
amplitude exceeds $0.1$\%.} 
  \label{Fig::Fe13_First}
\end{figure*} 
\subsection{Simplifying assumptions}
\label{sec::FormulationSub_Assumptions}
For the calculations presented in this work, we made several assumptions that considerably reduced their 
computational cost. First, P-CORONA allows for the option of imposing the Hanle saturation limit in the SEE. 
As noted in Section~\ref{sec::Intro}, this assumption can safely be made for the forbidden lines considered in this work. 
In addition, because the considered input radiation field corresponds to the photospheric continuum, it is essentially flat 
in the spectral vicinity of each of the considered lines. As such, the effects of Doppler brightening or dimming should 
not be expected to play any role, and their influence can be neglected in the SEE, allowing for a further decrease in 
computational cost.  

The atomic models considered for the calculations were obtained from the CHIANTI database %(version 10.0; see \citep{delZannaYoung20})
\citep[version 10.1; see][]{Dere+23}, but the total number of levels had to be limited for {computational} feasibility. 
As shown in \cite{SchadDima20}, for the four considered lines, the relative difference between the 
emission obtained from CHIANTI by considering the largest possible atomic models and 
by considering only the lowest $100$ levels does not exceed $5\%$, even within the electron density ranges for which the discrepancy is the largest 
\citep[but see also the recently published work by][which presents a reduced 55 level model for modeling the Fe~{\sc{xiii}} infrared forbidden lines]{DelZannaSupriya25}. 
In the present work, we applied P-CORONA to the \gls*{cbp} model, comparing the $\overline{I\,}$, $\overline{|V|\,}/\overline{I\,}$, 
and $P_L$ signals obtained when changing the number of levels considered in the corresponding 
atomic models, taking the $200$ level case as a benchmark. 
Because of the high computational demands of such tests, we carried out the required calculations 
by first reducing the spatial resolution of the \gls*{cbp} model from $512^3$ points to $151^3$ through linear interpolation. 
Compared to the benchmark, the atomic models with $31$ levels yield lower $\overline{I\,}$ values, 
but by at most $5\%$ for Fe$5303$, and by less that $3\%$ for the other considered lines. 
The $\overline{|V|\,}/\overline{I\,}$ signals vary by less than $1\%$ for any of the considered 
lines when taking only $31$ levels, compared to the benchmark case. 
Although the relative variations of $P_L$ in the $31$ level case are more significant, with local increases of 
over $20\%$ with respect to the benchmark being found for Fe$10747$, such variations are found precisely where the linear polarization amplitude is lowest. 
Indeed, we verified that such considerations do not affect the qualitative results shown in 
the following sections.
Thus, the synthetic intensity and polarization signals shown in the following sections were calculated by taking only the lowest 
$31$ levels for each atomic model. 

\section{Synthesis at a typical instant in the CBP evolution}
\label{sec::Stokes}

We first focus on the results of the P-CORONA synthesis for Fe$10747$, carried out for {the} \gls*{cbp} model described 
in Section~\ref{sec::FormulSubAtmosSSubCBP} (corresponding to an instant $201.7$ minutes from the start of the simulation). 
{The fan-spine magnetic topology of this model is illustrated in the left panel of Figure~\ref{Fig::Fe13_First}. 
The image shows a number of magnetic field lines (in red) traced from the neighborhood of a null point 
that delineate the fan surface (field lines going toward the horizontal boundaries of the domain) 
and the upper and lower spine (field lines that go toward the top and bottom of the domain). The} 
 wavelength-integrated intensity $\overline{I\,}$ {(upper right panel)}, 
$\overline{|V|} / \overline{I\,}$ (lower central panel), and $P_L$ {(lower right panel)} {calculated with P-CORONA are shown in the same figure}.
For reference, we also show the intensity mimicking a limb observation by SDO/AIA in the $193$~\AA\ channel (upper central panel), 
computed with CHIANTI \citep{Dere+23}. 
For this panel, we used the spatial coordinates taken directly from the numerical \gls*{cbp} model, rather than those 
corresponding to the input for P-CORONA (as in the rest of the figures shown in this work). The two coordinate systems are 
related through a $90^\circ$ rotation with respect to the $\mathrm{z}$ axis and a unit conversion to $R_\odot$, corresponding to $695.7$~Mm, 
while also considering that $\mathrm{z}=0$~Mm corresponds to $Z = R_\odot$.  

\begin{figure*}[!h]
\centering
\includegraphics[width = 0.94\textwidth]{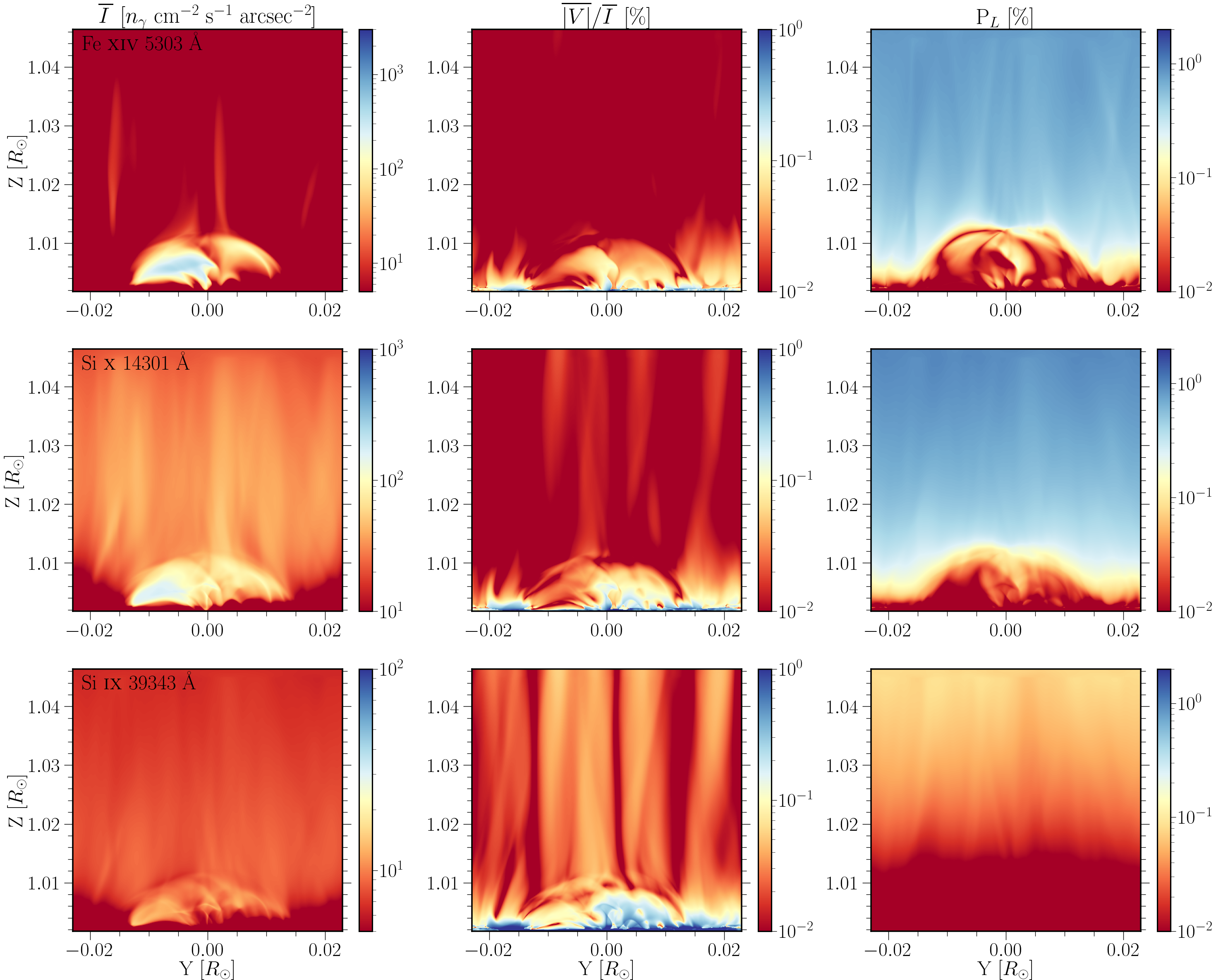}
\caption{Wavelength-integrated profiles obtained from P-CORONA calculations for the \gls*{cbp} model at $201.7$ minutes from 
 the start of the simulation.  
 \textit{Left, {central}, and right columns} show the quantities 
 $\overline{I\,}$, $\overline{|V|}/\overline{I\,}$, and $P_L$, respectively.
 \textit{Upper, middle, and lower rows} correspond to quantities obtained from the synthesis for 
 Fe$5303$, Si$14301$, and Si$39343$ lines, respectively.} 
  \label{Fig::Multilines_First}
\end{figure*} 
For the $\overline{I\,}$ signal of Fe$10747$, {there is} a clear enhancement within the \gls*{cbp} region, 
which is mostly contained less than $0.01\,R_\odot$ from the base of the corona. 
In the considered \gls*{cbp} model, the temperature and electron density are appreciably higher {underneath} 
the fan surface, leading to enhanced emission {and thus} the observed brightening. 
Within the \gls*{cbp} region, the $\overline{|V|}/\overline{I\,}$ signals 
reach values well above $0.1\%$, 
and could thus {be} expected to be detectable with \textcolor{black}{future} coronagraphs whose polarimetric sensitivity
and resolution is comparable to that of Cryo-NIRSP/DKIST but which can observe closer to the base of the corona. 
Interestingly, we verified that, throughout the \gls*{pos}, the 
$\overline{|V|}/\overline{I\,}$ signals are in good agreement with the ratio of the maximum $V$ over the maximum $I$  
in the spectral vicinity of the line (i.e., $|V_\mathrm{max}|/I_\mathrm{max}$). 

\begin{figure*}[!h]
\centering
\includegraphics[width = 0.94\textwidth]{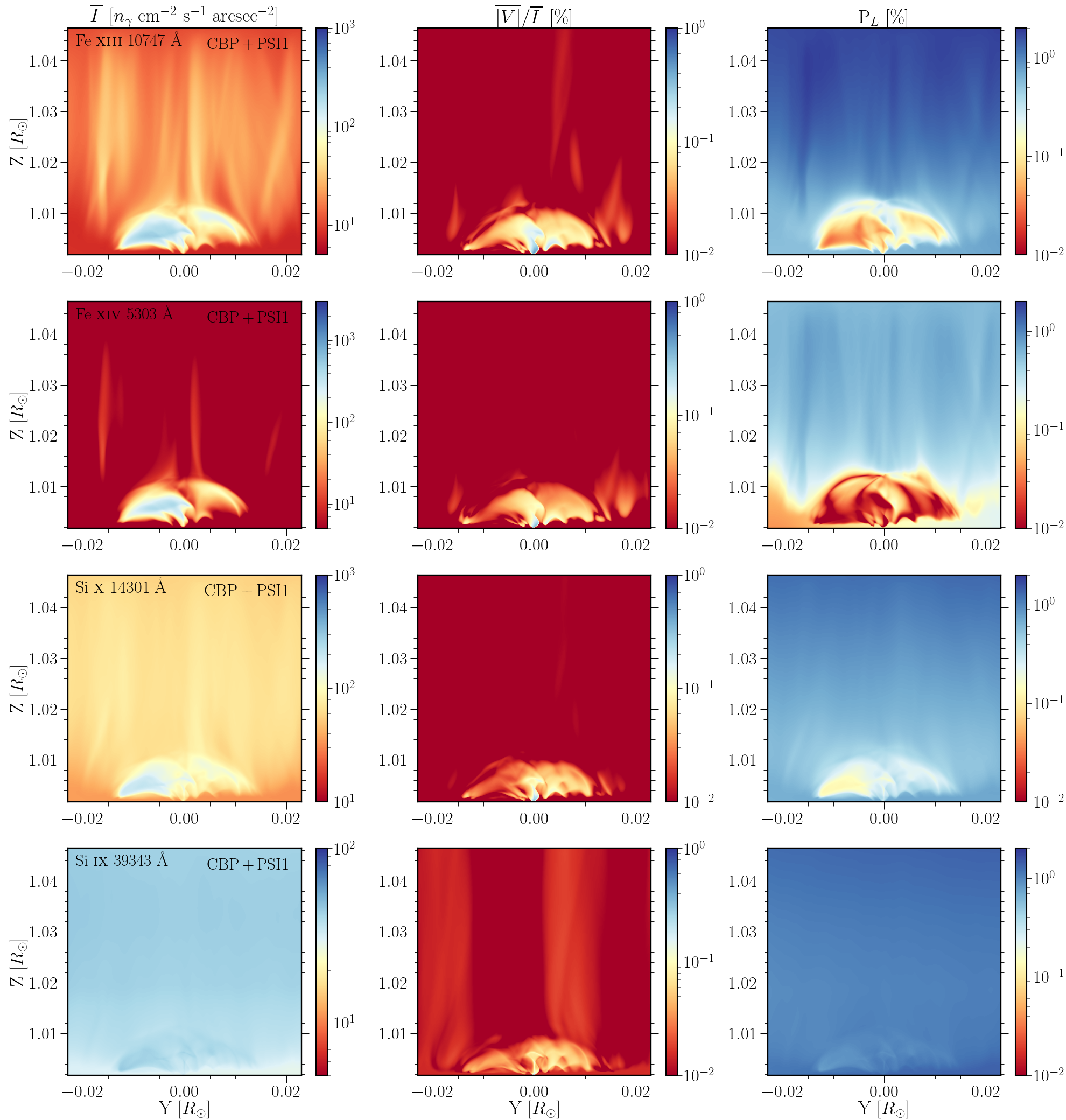}
\caption{Synthetic signals obtained by P-CORONA, for the radiation emitted by both the \gls*{cbp} and model PSI1 (see text), 
representative of the surrounding coronal material along the line of sight. 
\textit{Left, {central}, and right columns} show the quantities $\overline{I\,}$, $\overline{|V|}/\overline{I\,}$, 
and $P_L$, respectively. From top to bottom, the rows correspond to the quantities obtained from the synthesis for the Fe$10747$,
 Fe$5303$, Si$14301$, and Si$39343$ lines, respectively.}
 \label{Fig::Multi_Degrad_PS1}
\end{figure*}

\begin{figure*}[!h]
\centering
\includegraphics[width = 0.94\textwidth]{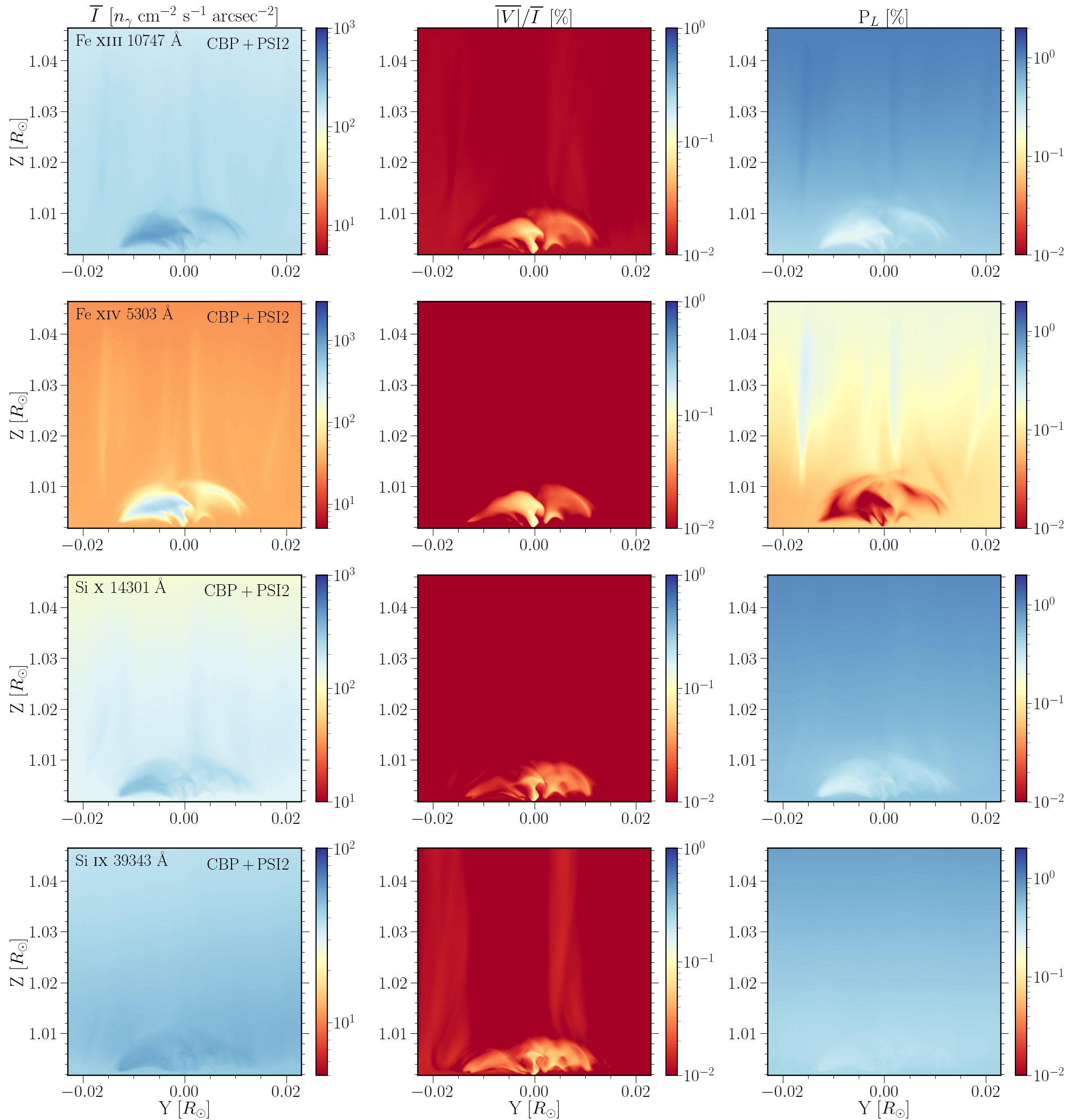}
\caption{Synthetic signals obtained by P-CORONA, for the radiation emitted by both 
the \gls*{cbp} and model PSI2 (see text), representative of the surrounding coronal material along the line of sight. 
\textit{Left, {central}, and right columns} show the quantities $\overline{I\,}$, $\overline{|V|}/\overline{I\,}$, 
and $P_L$, respectively. From top to bottom, the rows correspond to the quantities obtained from the synthesis for the Fe$10747$,
 Fe$5303$, Si$14301$, and Si$39343$ lines, respectively.}
 \label{Fig::Multi_Degrad_PS2}
\end{figure*}
The linear polarization, on the other hand, reaches its largest amplitudes outside the 
\gls*{cbp}. At $0.02\,R_{\odot}$ above the base of the corona, the $P_L$ amplitude is on the order of
$1\%$, but it is strongly suppressed within the \gls*{cbp} region, where %its amplitudes fall
it drops below $0.1\%$ and often even below $0.01\%$. We recall that both collisional and radiative processes contribute
to the excitation of the upper level of the transition but, assuming isotropic collisions, only the absorption 
of anisotropic radiation leads to atomic polarization in the upper level of the line transition and, thus, to linear polarization. 
The aforementioned enhanced electron density within the \gls*{cbp} region does not only induce an increased 
intensity brightening; it additionally causes a higher rate of collisional excitations, leading to cascades that further populate 
the upper level, and a higher rate of collisional depolarization. 
As a result, the atomic polarization of the upper level decreases and, 
consequently, $P_L$ is also reduced 
\citep[for a discussion on the impact of collisions on the atomic alignment in high-density 
regions of the corona, see][]{SchadDima20}. 
Indeed, we verified that the linear polarization amplitude would be even lower within the 
\gls*{cbp} (decreasing further by up to $20\%$) if $200$ levels were taken for the atomic model instead of $31$. 

The calculations for the other three lines investigated in this work are shown in Figure~\ref{Fig::Multilines_First}. 
The four lines can be ordered according to their intensity within the \gls*{cbp}, from brightest to darkest, as 
Fe$5303$, Fe$10747$, Si$14301$, and Si$39343$ 
(we note that, in the figures, the intensity scale is adjusted for each considered line).   
We find the same ordering, from highest to lowest, for their peak response temperature and for the Einstein coefficients for 
spontaneous emission (see Table~\ref{Tab::Table1}). 
On the other hand, outside the \gls*{cbp} region, the highest intensity is found for Si$14301$, 
which has a peak emission {around} $T = 10^{6.15}$~K {or} $1.4$~MK, whereas significantly lower intensities are found for the other lines. 
The contrast between the intensity within the \gls*{cbp} and outside it is particularly low for the 
Si$39343$ line, whose peak emission falls at {around} $T = 10^{6.05}$~K, {or} $1.1$~MK.  

Regarding the $\overline{|V|}/\overline{I\,}$ signals in the \gls*{cbp}, we find the 
highest amplitudes for Si$39343$, with intermediate amplitudes for Fe$10747$ and Si$14301$, and the lowest for Fe$5303$. 
However, even for the latter line, such signals often approach $0.1\%$. The differences between the lines could be 
expected from the fact that the circular polarization patterns arise from the Zeeman effect, whose amplitude scales 
with the wavelength of the considered line. 
As for the linear polarization, we find the $P_L$ amplitudes to be highest for 
Fe$10747$, {followed by} Si$14301$, Fe$5303$, and {lowest for} Si$39343$
\citep[in agreement {with the results of}][{which correspond to calculations at $0.07\,R_\odot$ above 
the solar limb using an idealized, spherically symmetric, hydrostatic, and isothermal coronal model}]{Judge+06}. 
For all lines, {the} $P_L$ is strongly suppressed within the \gls*{cbp}, for the reasons explained above. 
Thus, we consider the linear polarization to be of limited utility for diagnostics of the magnetic fields in 
the low-corona \gls*{cbp} regions. Nevertheless, for completeness, we show in Appendix~\ref{sec::AppPolang} 
the linear polarization angles (defined as in Equation~\ref{Eq::PolAng}) associated with the calculations presented 
in this section. 

\section{Coronal material along the line of sight}
\label{sec::LoSMat}
We now analyze how the intensity and polarization patterns are modified in the more realistic scenario of accounting for the 
coronal material {lying} along the \gls*{los}, beyond the spatial domain of the \gls*{cbp} model 
(which extends little more than $30$~Mm, or $0.045~R_\odot$, in any spatial direction). We consider 
the impact of models PSI1 and PSI2 introduced in Section~\ref{sec::FormulSubAtmosSSubPS}, which 
extend from $-3\,R_\odot$ to $3\,R_\odot$ in the \gls*{los} direction, for the four forbidden lines considered in this work. 
Model PSI1 can be considered to represent a coronal hole, 
and its temperature is {slightly} lower than that of model PSI2:
their average temperature {is} $1.07$~MK and $1.15$~MK, respectively, 
whereas their average electron {density is} $2.8 \cdot 10^7$ cm$^{-3}$ and $2.9 \cdot 10^7$ cm$^{-3}$, respectively. 
Particularly for the case of the lines with a high peak response temperature, 
we verified that PSI1 presents a substantially lower emission %-- integrated along the LoS -- \fmicom{remove the hyphens "--"}
integrated along the \gls*{los} 
than model PSI2. For instance, for the Fe$10747$ line, the integrated emission is more than one order of magnitude lower, 
which is in rough agreement 
with the difference in the contribution functions of the 
line at $1.07$ and $1.15$~MK \citep[e.g., Figure~2 of][]{SchadDima20}. To compare the role played
by temperature and electron density in the resulting emission, we verified that artificially increasing the temperature by $25\%$ at all spatial points of the PSI1 model increases the resulting emission by roughly a factor of $10$, whereas the same increase in density leads to more moderate increase in emission, 
of roughly $40\%$. 

For the four lines of interest, the $\overline{I\,}$,  $\overline{|V|}/\overline{I\,}$, and $P_L$ signals
obtained when considering the \gls*{cbp} model together with either PSI1 or PSI2 are shown in 
Figures \ref{Fig::Multi_Degrad_PS1} and \ref{Fig::Multi_Degrad_PS2}, respectively. 
We begin the discussion by focusing on the impact of the PSI models on the Fe$10747$ line, which corresponds to the upper row of the figures, 
comparing them to the calculations considering the \gls*{cbp} model alone, shown in Figure~\ref{Fig::Fe13_First}.  
In intensity, we find that the contribution from PSI1 does not appreciably impact the signal from the \gls*{cbp} region, 
although it is certainly noticeable outside it. On the other hand, the contribution 
from model PSI2 has a strong impact on the intensity signal both outside the \gls*{cbp} and within 
it. We find a similar impact on the $\overline{|V|}/\overline{I\,}$ signals; inside the \gls*{cbp} region, 
the circular polarization amplitude is barely affected by PSI1, but it is greatly reduced when including 
instead the contribution from PSI2. The magnetic fields in PSI1 and PSI2 
have strengths of $\sim\!\!1$~G, which is substantially lower than the ones found in the high-density regions of the \gls*{cbp} model. 
Because of their relatively low magnetic field strength, when the PSI models contribute significantly to the amplitude of the 
intensity profile, they do not proportionally enhance the amplitude of the Stokes $V$ profile, which is produced by the Zeeman effect.
Thus, a substantial contribution from the PSI models to $\overline{I\,}$ results in a net decrease of $\overline{|V|}/\overline{I\,}$. 

The linear polarization signals of the Fe$10747$ line are far more vulnerable to the contribution 
from the surrounding material, which appreciably enhances the amplitude of the resulting $P_L$
signals in the \gls*{cbp} region, even for the PSI1 model. 
The electron density {of such material is lower} than {that of the plasma underneath} the \gls*{cbp} 
fan surface. Because of this, even if the emission is comparatively low, the relative contribution from scattering processes
increases, leading to a larger linear polarization amplitude. 
Moreover, because the surrounding material is farther from the base of the corona than
the \gls*{cbp}, the radiation reaching it from the photosphere is more anisotropic, further contributing to %an enhancement of
the scattering polarization amplitude. For instance, when considering PSI2, we verified numerically that most of the contribution to the overall 
$P_L$ comes from the points along the \gls*{los} between roughly $0.2$ and $0.4$ solar radii from the base of the corona, 
where the anisotropy of the radiation increases substantially, compared to its value below $0.1$ 
solar radii \citep[e.g.,][]{Li+17}. 

By comparing Figures~\ref{Fig::Multilines_First}, \ref{Fig::Multi_Degrad_PS1}, and \ref{Fig::Multi_Degrad_PS2}, we 
can evaluate the impact of the PSI models on the other forbidden lines under investigation. 
Regarding the intensity, we observe that the coronal material from these models has the smallest impact on the signals 
of Fe$5303$, whose peak response temperature is the highest, at $\sim\!\!10^{6.3}$~K (see Table~\ref{Tab::Table1}).  
The contribution from PSI1 has no appreciable impact on the intensity, 
even outside the \gls*{cbp}. When considering the contribution from PSI2, we do find an 
enhancement outside the \gls*{cbp}, although the signal within it remains largely unaffected. 
As for Si$14301$, the contribution from {PSI1} clearly impacts the intensity both {inside and} outside 
the \gls*{cbp}, {slightly} more so than for the case of Fe$10747$. 
Model %B 
PSI2 has a far stronger impact, to the point that the \gls*{cbp} is {barely} distinguishable from the background coronal intensity. 
For the case of Si$39343$, which has {the} lowest peak response temperature, the \gls*{cbp} signal is completely drowned out by the contribution 
of the surrounding coronal material, whether model 
PSI1 or PSI2 is considered. 

For the cases presented in this section, we find that the $\overline{|V|} / \overline{I\,}$ 
signals within the \gls*{cbp} region are generally impacted by the contribution from the PSI models 
to a similar degree as the $\overline{I\,}$ signals. 
We expect that, if a \gls*{cbp} can clearly be observed in the intensity signals of a given line, 
the detected circular polarization patterns can be mainly attributed to the emission from the \gls*{cbp} region.  

Regarding the linear polarization fraction, the findings for the Si$14301$ and Si$39343$ lines 
are similar to those for Fe$10747$: the $P_L$ signals in the 
\gls*{cbp} are strongly affected even in cases where the $\overline{I\,}$ signals are not. 
For the Fe$5303$ line, on the other hand, the contribution from %A or B is so low
PSI1 or PSI2 is so low, compared to that from the \gls*{cbp}, that the $P_L$ signals are only slightly affected even when considering PSI2, 
{and thus their amplitude} remains small. These results reinforce the conclusion that the 
linear polarization signals {of these lines} are not suitable diagnostics for the magnetic field in the \gls*{cbp} region.
% % 

\begin{figure*}[!h]
 \centering
 \includegraphics[width = 0.94\textwidth]{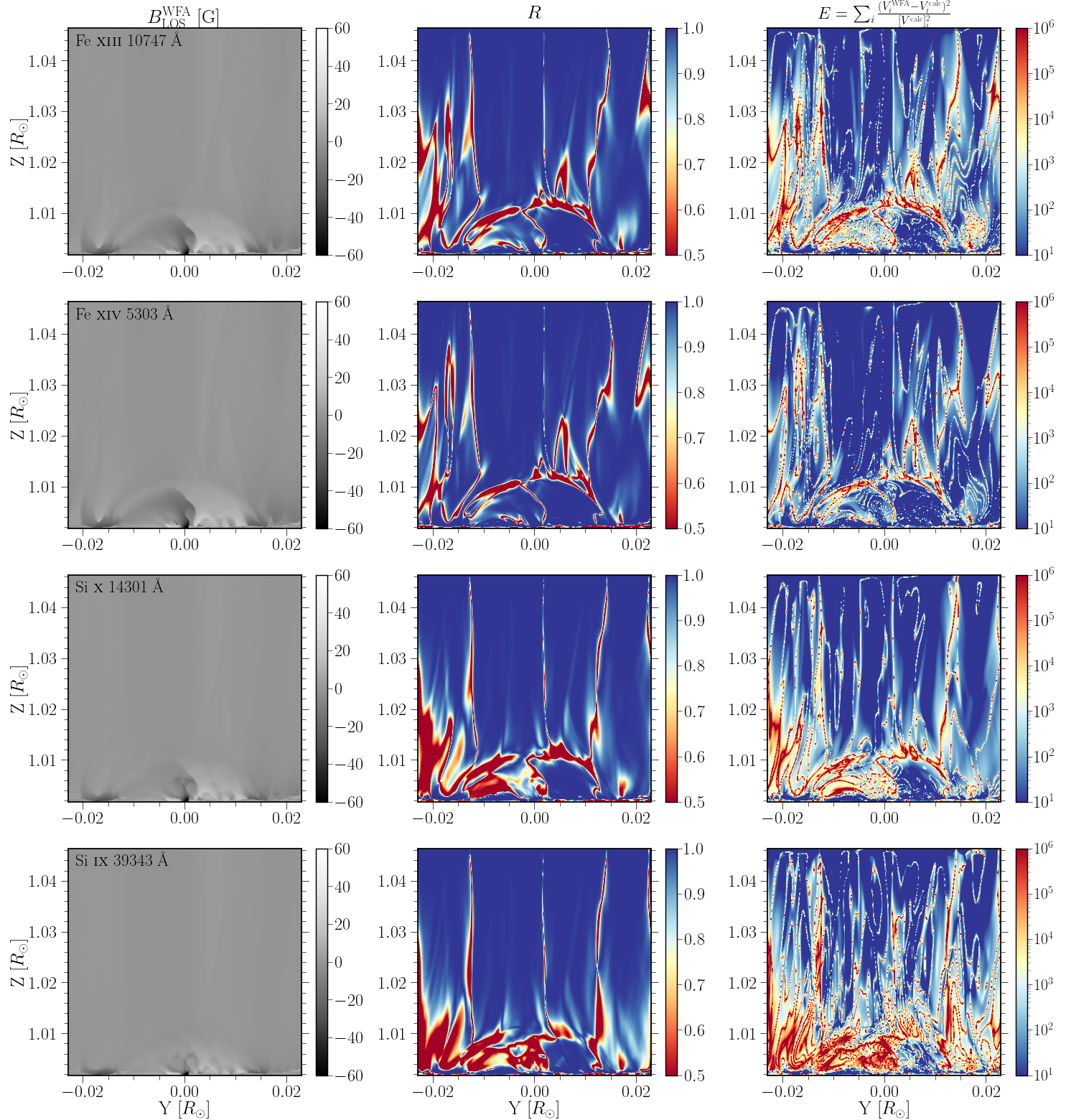}
 \caption{Quantities resulting from application of the WFA. From top to bottom, the four rows show the quantities for the Fe$10747$, Fe$5303$, Si$14301$, and Si$39343$ lines. 
 \textit{Left column:} Line-of-sight magnetic fields, 
 inferred through the application of the WFA to the synthetic profiles obtained from P-CORONA (see text). 
 \textit{Central column:} Pearson coefficient for the correlation between the $V(\lambda)$ and $\partial I(\lambda)/\partial\lambda$ synthetic profiles, at each point on the \gls*{pos}. \textit{Right column:} Square relative error between the $V(\lambda)$ profiles resulting from the synthesis and 
  from the application of the WFA.}
  \label{Fig::Multi_WFA}
\end{figure*}

\section{Applicability of the weak field approximation}
\label{sec::WFA}
In the present section, we analyze the value of the circular polarization signals 
for inferences of the longitudinal component of the magnetic {field} within the regions that emit in the considered forbidden lines. 
The circular polarization patterns presented above {account} for the energy shift between different sublevels 
of the same $J$ level, induced by the magnetic field (i.e., the Zeeman effect). %
The \gls*{los} component of the magnetic field can be inferred {by} making use of the \gls*{wfa}, 
through which the circular polarization can be related to the wavelength derivative of the intensity profile according to 
\citep[e.g., ][]{TrujilloBuenodelPinoAleman22} 
\begin{equation}
      V(\lambda) = - 4.67 \times 10^{-13} \, g_{\mathrm{eff}} \, \lambda_0^2 \, 
      B_{\mathrm{LOS}} \, \biggl(\frac{\partial I(\lambda)}{\partial \lambda} \biggr) \, ,
    \label{Eq::WFA}
\end{equation} 

where $g_{\mathrm{eff}}$ is the (dimensionless) effective Land\'e factor of the transition, %$e$ is the electron charge, $m_e$ is the mass of the electron, 
$\lambda_0$ is the wavelength of the considered line transition {in \AA}, and $B_{\mathrm{LOS}}$ is the \gls*{los} 
component of the magnetic field {in G}. The partial derivative with respect to $\lambda$ {is also given} in \AA\ units.
As noted by \cite{CasiniJudge99}, 
this relation should also contain a correction factor due to the atomic alignment of the 
upper level of the line. However, within the \gls*{cbp}, {the} atomic alignment should be very small, as a consequence of 
the relatively high density and predominance of collisional processes (discussed in Section~\ref{sec::Stokes}). 
On this basis, the corresponding correction factor is neglected in the present work. 
The application of the \gls*{wfa} hinges on a few assumptions, 
notably that the magnetic splitting must be much smaller than the Doppler width of the lines 
(which is {valid} for the conditions for the coronal lines considered in this work). 
It also requires the longitudinal component of the magnetic field to 
be constant within the emitting medium. Such {an} assumption is not guaranteed 
in the considered atmospheric model, and its influence 
{must be examined}. 

\begin{figure*}[!h]
 \centering
 \includegraphics[width = 0.94\textwidth]{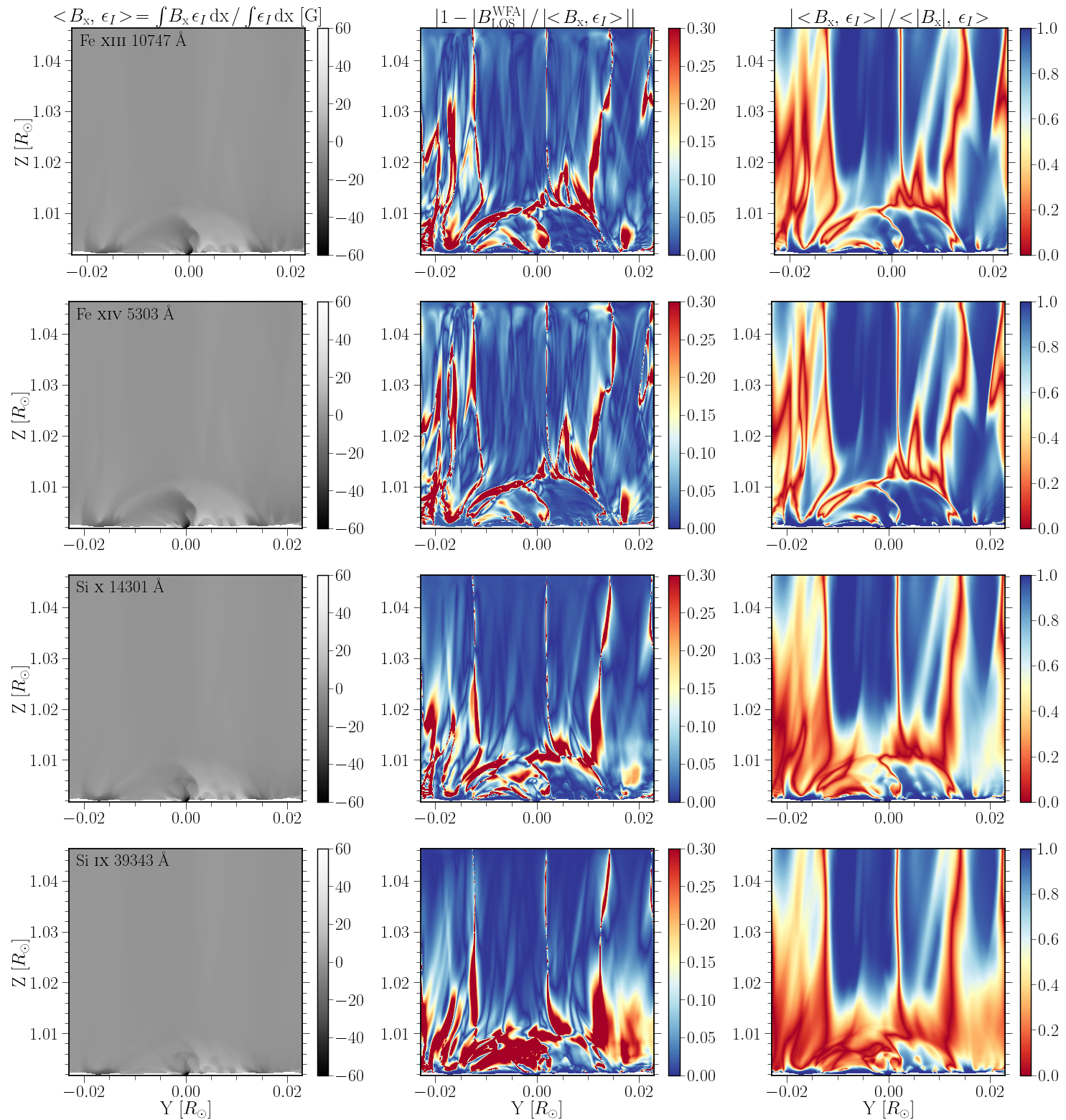} 
 \caption{
 Quantities for the WFA analysis, taken from the values of the numerical \gls*{cbp} model. 
 From top to bottom, the four rows show the corresponding quantities when considering the Fe$10747$, Fe$5303$, Si$14301$, and Si$39343$ lines.
 \textit{Left column:} Spatial average along the \gls*{los} direction of $B_\mathrm{LOS}$, weighted by the emission of the considered
 line as computed by CHIANTI. \textit{Central column:} Departure from unity of the ratio between the B$_\mathrm{LOS}$ inferred from the \gls*{wfa} and 
 the average $B_\mathrm{LOS}$ of the atmospheric model, weighted by the emission. 
 \textit{Right column}: Measure of the cancellation of $B_\mathrm{LOS}$ along the 
 \gls*{los} 
 {given as the ratio of the integral quantities $|\!\!<\!B_\mathrm{x}, \epsilon_I \!>\!\!|$ over 
 $<\!|B_\mathrm{x}|, \epsilon_I \!>$} (see text). 
 }
 \label{Fig::Multi_Mod_BLOS}
\end{figure*}

Given the $I(\lambda)$ and $V(\lambda)$ profiles, we can use Equation~\ref{Eq::WFA} to infer the average \gls*{los} 
component of the magnetic field  (i.e., $B^{\mathrm{WFA}}_\mathrm{LOS}$) through least-squares minimization. 
{We begin the discussion focusing on the Fe$10747$ line {(Figure~\ref{Fig::Multi_WFA}, upper row)}}. 
The upper left panel shows the magnetic {field} inferred when applying the \gls*{wfa} to 
the profiles obtained {for the \gls*{cbp} model using} P-CORONA 
(corresponding to the calculations whose resulting wavelength-integrated signals are shown 
in Figure~\ref{Fig::Fe13_First}). 
The least-squares minimization was computed taking the wavelength grid points provided by P-CORONA 
within a range of $6$~\AA , centered at the wavelength of the transition. 
Within the \gls*{cbp}, we find both positive and negative longitudinal magnetic fields 
(i.e., pointing towards and away from the observer, respectively), 
{with strength exceeding $50$~G, which is significantly higher than outside of it}. 
We note that this approach gives information on the magnetic field within the regions from which the line's radiation is emitted. 

The circular polarization patterns shown, for instance, in Figure~\ref{Fig::Fe13_First}, do not originate from a localized source. 
Instead, they are emitted from an extended spatial range, along which the thermodynamic properties
including $B_{LOS}$ may vary, thus compromising the applicability of the \gls*{wfa}. 
As a result, the $V(\lambda)$ profile may not be proportional to $\partial I(\lambda) / \partial \lambda$. The upper central panel of 
Figure~\ref{Fig::Multi_WFA} shows, for each pixel on the PoS, the Pearson correlation coefficient between the two 
profiles of the Fe$10747$ line, taking the same wavelength range as for the least-squares minimization. 
A value of $1$ would thus indicate perfect correlation and zero indicates that they are completely uncorrelated quantities. 
{Within much of the CBP},  
the coefficient is close to unity, but it 
tends to fall below $0.6$ (indicating a low correlation) close {to} the regions on the \gls*{pos} where there is a change in sign 
in $B_\mathrm{LOS}$ or it otherwise becomes small. In addition, we computed the square error as
\begin{equation*}
    \label{Eq::Error}
    E = \sum^{N_\lambda}_i \frac{(V_i^{\mathrm{WFA}} - V_i^{\mathrm{calc}})^2}{(V_i^\mathrm{calc})^2} \, ,
\end{equation*}
where $V^{\mathrm{calc}}$ is the circular polarization computed using P-CORONA and $V^{\mathrm{WFA}}$ is 
obtained according to Equation~\eqref{Eq::WFA}, taking the derivative of the intensity 
from the P-CORONA calculation and using the value of $B_{\mathrm{LOS}}$ found through a least-squares minimization. 
The sum over $i$ encompasses the same spectral grid points as for the least-squares minimization. 
We find that, within the \gls*{cbp}, the square error is largest where the Pearson coefficient becomes {smallest}.  
Again, this coincides with the regions on the PoS where the inferred $B_\mathrm{LOS}$ is small. 

As a further test of the suitability of the inference of the \gls*{los} magnetic field using the \gls*{wfa},  
we calculated the average $B_{\mathrm{LOS}}$ in the \gls*{los} direction directly from the 
{values of the numerical} \gls*{cbp} model. 
We performed the averages by weighting the magnetic field at each spatial point by the emissivity in 
the considered spectral line $\epsilon_I$, computed as {a} function of temperature, electron density and 
distance from the solar surface using CHIANTI {as follows:} 
\begin{equation}\label{eq:weighted_average}
    <\!B_\mathrm{x}, \epsilon_I \!>\, = \frac{\int B_\mathrm{x} \, \epsilon_I \, \mathrm{d}\mathrm{x}}{\int \epsilon_I \, \mathrm{d}\mathrm{x}} \, ,
\end{equation}
where we recall that $\mathrm{x}$ is the \gls*{los} direction and thus $B_\mathrm{x}$ is equivalent to $B_{\mathrm{LOS}}$ 
in the \gls*{cbp} model. 
Focusing first on the Fe10747 line, the resulting average is shown first in the upper 
left panel of Figure~\ref{Fig::Multi_Mod_BLOS}. 
Visual comparison between this quantity and the \gls*{los} component of the magnetic field as inferred from the \gls*{wfa} 
(see the upper left panel of Figure~\ref{Fig::Multi_WFA}) shows a generally good agreement. 
Such a comparison is shown directly in the {upper central} panel of Figure~\ref{Fig::Multi_Mod_BLOS}, in terms of 
the departure from unity of the ratio between the absolute value of the two quantities. In most of the \gls*{cbp} region, 
the two quantities deviate by less than $5\%$, although discrepancies as large as $30\%$ can be found, coinciding with the 
regions with a low correlation coefficient between $V^{\mathrm{calc}}$ and $V^{\mathrm{WFA}}$. 
This generally good agreement is consistent with the findings of \cite{SchadDima20}, who 
performed calculations using the PyCELP code and a MURaM simulation from \cite{Rempel17}. 
That model is representative of a {relatively simple}  
bipolar active region and yielded substantially weaker longitudinal magnetic fields. 
As suggested by those authors, the remarkable agreement between the two quantities is due to the fact that the 
\gls*{wfa} gives more weight to the regions along the \gls*{los} where most of the emission occurs in the considered line. 
\begin{figure}[!h]
    \centering
    \includegraphics[width=0.31\textwidth]{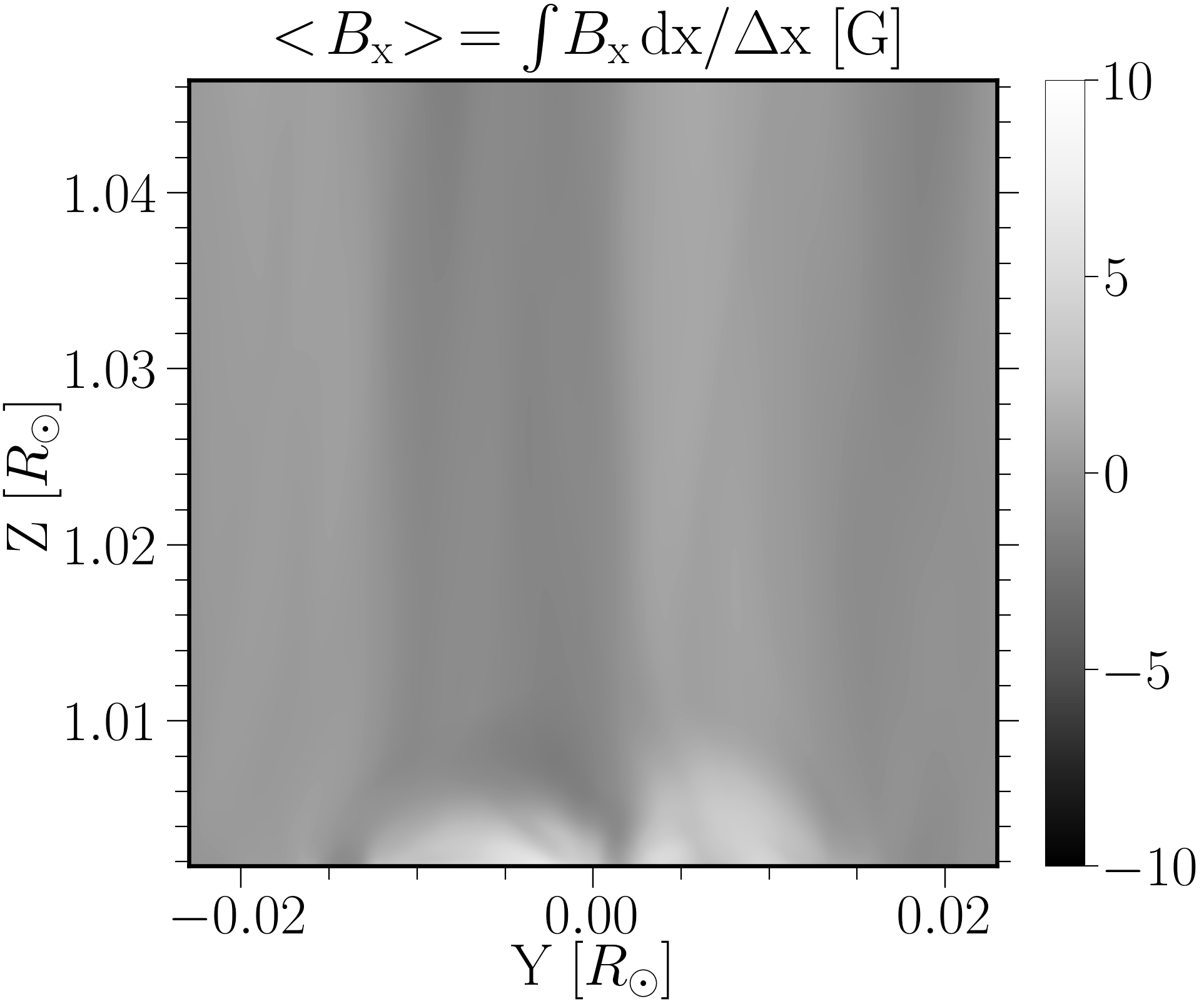}
    \caption{Spatial average, along the \gls*{los} direction, of the \gls*{los} component 
    of the magnetic field, $B_{\mathrm{LOS}}$, as taken from the \textit{Bifrost} \gls*{cbp} model discussed in the main text.}
    \label{fig::AvModBLOS}
\end{figure}

To further illustrate this point, we first show in Figure~\ref{fig::AvModBLOS} the quantity 
$<\!\!B_\mathrm{x}\!\!>\, = \! \int \!B_\mathrm{x} \, \mathrm{d}\mathrm{x}$,  
whose value is very different from that of $<\!B_\mathrm{x}, \epsilon_I \!>$. The former is predominantly {positive} within the \gls*{cbp}, 
whereas the latter shows distinct regions with positive and negative sign. The values of 
$<\!\!B_\mathrm{x}\!\!>$ are substantially lower than those of $<\!B_\mathrm{x}, \epsilon_I \!>$, 
suggesting that relatively strong magnetic fields are present in the high-temperature regions 
of the \gls*{cbp} where the emission occurs. 
In addition, we computed the ratio $\bigl|\int B_\mathrm{x} \,\epsilon_I \,\mathrm{d}\mathrm{x} \bigr| \, \big/ \int |B_\mathrm{x}| \,\,\epsilon_I \,\mathrm{d}\mathrm{x}$, or, 
using the symbol defined in {Equation~}(\ref{eq:weighted_average}),  $|\!\!<\!\!B_\mathrm{x}, \epsilon_I \!\!>\!\!|$ over 
 $<\!\!|B_\mathrm{x}|, \epsilon_I \!\!>$. 
This quantity gives a measure of the magnetic field cancellations along the \gls*{los}, while 
giving more weight to the spatial regions where the emissivity is strongest.  
If the magnetic field had the same sign along the integration domain, then the numerator and the denominator 
would have the same value and the ratio would be unity. 
Conversely, if the sign were to change in such a manner 
that the emission-weighted magnetic fields perfectly cancel along the \gls*{los}, 
the resulting quantity would be zero. 
This ratio, for the same 
\gls*{cbp} model and considering 
the Fe$10747$ line, is shown in the upper {right} panel of Figure~\ref{Fig::Multi_Mod_BLOS}. 
The regions with the greatest cancellation along the \gls*{los} coincide with those with the highest 
discrepancy between the magnetic fields inferred through the \gls*{wfa} and those resulting from the 
average over the model itself, and where the correlation coefficients are lowest. 

We now present the same analysis for the other lines, showing 
 the corresponding quantities in the second, third, and fourth rows of 
Figures~\ref{Fig::Multi_WFA} and \ref{Fig::Multi_Mod_BLOS} 
for Fe$5303$, Si$14301$ and Si$39343$, respectively.  
When applying the least-squares minimization for the \gls*{wfa}, we took a $6$~\AA\ spectral interval 
for the Fe$5303$ and a $10$~\AA\ interval 
for both Si$14301$ and Si$39434$. 
For Fe$5303$, we obtain $B^\mathrm{WFA}_\mathrm{LOS}$ values that are similar to 
those obtained when considering the Fe$10747$ line. 
Indeed, the results for the associated quantities are also very similar; 
the correlation coefficient (central column of Figure~\ref{Fig::Multi_WFA}) and the departure from unity of the ratio between 
$B^\mathrm{WFA}_\mathrm{LOS}$ and $<\!B_\mathrm{x}, \epsilon_I\!>$ 
(central column of Figure~\ref{Fig::Multi_Mod_BLOS})  
both show their lowest (i.e., worst) values in the same regions on the \gls*{pos} 
where $B_{\mathrm{LOS}}$ has its lowest values. Of course, such regions also tend to coincide 
with the regions where the square error (right column of Figure~\ref{Fig::Multi_WFA}) 
is the largest and for which the strongest cancellations occur, as indicated by low values of 
$|\!<\!B_\mathrm{x}, \epsilon_I\!>\!| \,\bigr/ \!<\! |B_\mathrm{x}|,\, \epsilon_I\!>\!$ 
(right column of Figure~\ref{Fig::Multi_Mod_BLOS}). 

\begin{figure*}[!h]
 \centering 
  \includegraphics[width = 0.94\textwidth]{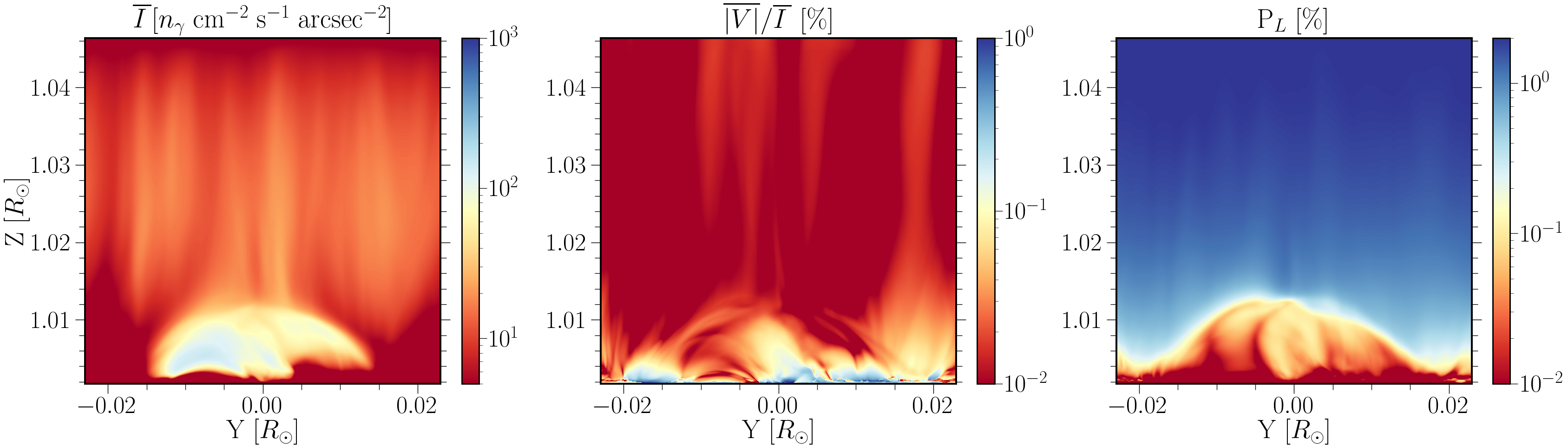}
 \caption{Time average for the Fe~{\sc{xiii}} line at $10747$~\AA , 
 resulting from syntheses taking the \gls*{cbp} model at instants of the simulation between 
 $185$ and $218.3$ minutes, with $100$-second cadence. 
 The panels correspond to the same quantities as in Figure~\ref{Fig::Fe13_First}.} 
  \label{Fig::Fe13_Stok_Evol}
\end{figure*}

On the other hand, the qualitative differences are more considerable for the results of the 
Si$14301$ and, even more so for the Si$39343$ line. There is a clear increase in the area on the \gls*{pos} 
where the discrepancy between $B_\mathrm{LOS}^\mathrm{WFA}$ and $<\!\!B_\mathrm{x}, \epsilon_I\!\!>$ is large ($>25\%$), where the 
Pearson coefficient falls below $0.6$, and where the square error $E$ is large. 
Again, such regions coincide with those where there are large cancellations of the magnetic field along the \gls*{los}. 
Significantly lower magnetic fields are found when applying the \gls*{wfa} to such lines. 
Because these lines have a lower peak response temperature, they are emitted over a more extended region within 
the \gls*{cbp} than in the case of the Fe$10747$ and Fe$5303$ lines. As a result, when considering the integration 
of $B_{\mathrm{LOS}}$ along the \gls*{los}, weighted by the emissivity, cancellations occur more frequently. 

{The investigations presented in this section were carried out considering the profiles obtained when considering 
only the \gls*{cbp} model. However, as noted in Sect.~\ref{sec::LoSMat}, when the contribution from the surrounding 
material significantly impacts the intensity signal in the \gls*{cbp}, it should be expected to also 
impact the ${|\overline{V}|} / {\overline{I\,}}$ signal. This, in turn, would cause the 
$B_{\mathrm{LOS}}$ inferred via the \gls*{wfa} to be underestimated. 
We evaluated the impact on the $B_{\mathrm{LOS}}^{\mathrm{WFA}}$ inferred within the 
\gls*{cbp} region, although the corresponding figures are not shown for the sake of brevity. 
In the case of Fe$5303$, the field strength is underestimated by less than 
$5\%$ within most of the \gls*{cbp} when considering PSI1 and by less than $20\%$ 
when considering PSI2. 
For Fe$10747$, the underestimation is generally less than $15\%$ when considering PSI1, 
but it approaches or exceeds $80\%$ in most of the \gls*{cbp} region for PSI2. 
For Si$14301$, the decrease in the 
inferred $B_{\mathrm{LOS}}$ is appreciably greater than for the case of Fe$10747$ for 
either of the PSI models. For Si$39343$, in agreement with what could be expected from the results 
shown in Figures~\ref{Fig::Multi_Degrad_PS1} and \ref{Fig::Multi_Degrad_PS2}, the contribution 
from the surrounding material strongly contaminates the \gls*{cbp} signal whether PSI1 or PSI2 is considered, 
rendering the 
\gls*{wfa} inapplicable. } 
 
We thus regard Fe$10747$ and Fe$5303$ as the most suitable lines for the application of the \gls*{wfa}.  
Moreover, the circular polarization amplitude, normalized to intensity, 
is significantly larger for Fe$10747$ than 
{for} Fe$5303$ within the \gls*{cbp} region, because of the former's longer wavelength. 
Because of this, {we} will focus on the Fe$10747$ line in the following section. 

\section{Evolution over time}
\label{sec::Timint}
{Even state-of-{the-art} coronagraphs such as Cryo-NIRSP/DKIST require exposure times of tens of minutes to achieve a \gls*{snr} 
that allows measuring $V/I$ signals in the Fe$10747$ line with amplitudes of around $0.1\%$ (e.g., \citealp{Schad+24}).} 
In the present section, we investigate how the signals of this line change when accounting for 
the evolution of the \gls*{cbp} and averaging over time spans of tens of minutes, rather than 
considering the signals at specific instants as was done in the previous sections. 
Thus, we consider the \gls*{cbp} model discussed in Section~\ref{sec::FormulationSub_Bifrost} at 
several instants between $t=185$ and $218.3$~min, 
corresponding to a total interval of $33.3$ minutes or $2000$ seconds. 
To obtain the time-averaged signals, we summed the Stokes profiles obtained with P-CORONA from the \gls*{cbp} model 
with a cadence of $100$~s and divided by the total time interval\footnote{We also 
considered averages with lower cadences, and found no change in the resulting quantities. 
We thus expect that calculations at higher cadences would not yield different results.}.  
The resulting average $\overline{I\,}$, $\overline{|V|}/\overline{I\,}$, and $P_L$ signals 
is shown in Figure~\ref{Fig::Fe13_Stok_Evol}. An animation showing the corresponding quantities 
for different instants in time increments of $100$ seconds can be found in the online material. 
 
The left panel of the figure shows the average $\overline{I\,}$ signals, in which one can see 
that the signatures are somewhat smeared due to the time evolution 
\textcolor{black}{compared to the results at a given instant 
(see Figure~\ref{Fig::Fe13_First})}, both within the \gls*{cbp} and outside it. 
Nevertheless, the \gls*{cbp} itself is still clearly appreciable, and is brighter than the surrounding 
region by about an order of magnitude. 
The $\overline{|V|}/\overline{I\,}$ image, shown in the central panel, is also somewhat smeared by the time evolution: 
there is an appreciable decrease both in the maximum amplitude and in the area within the \gls*{cbp} where the 
amplitude exceeds $0.1\%$. Regardless, the area where the amplitude is above $0.1\%$ remains considerable. 

As can be seen in the right panel of the figure, the amplitude of the linear polarization increases within the \gls*{cbp}, 
compared to the polarization 
at a single instant (see Figure~\ref{Fig::Fe13_First}). 
This is because there are instants during the evolution 
{in which}, for a given point on the \gls*{pos} within the \gls*{cbp} region, 
the density is comparatively low throughout the \gls*{los}, thus contributing to a higher $P_L$. 
However, for the case of Fe$10747$, this contribution would still be \textcolor{black}{drowned out} by the 
external coronal material {lying} on the \gls*{los}, as discussed in Section~\ref{sec::LoSMat}. 

\begin{figure*}[!h]
 \centering
  \includegraphics[width=0.94\textwidth]{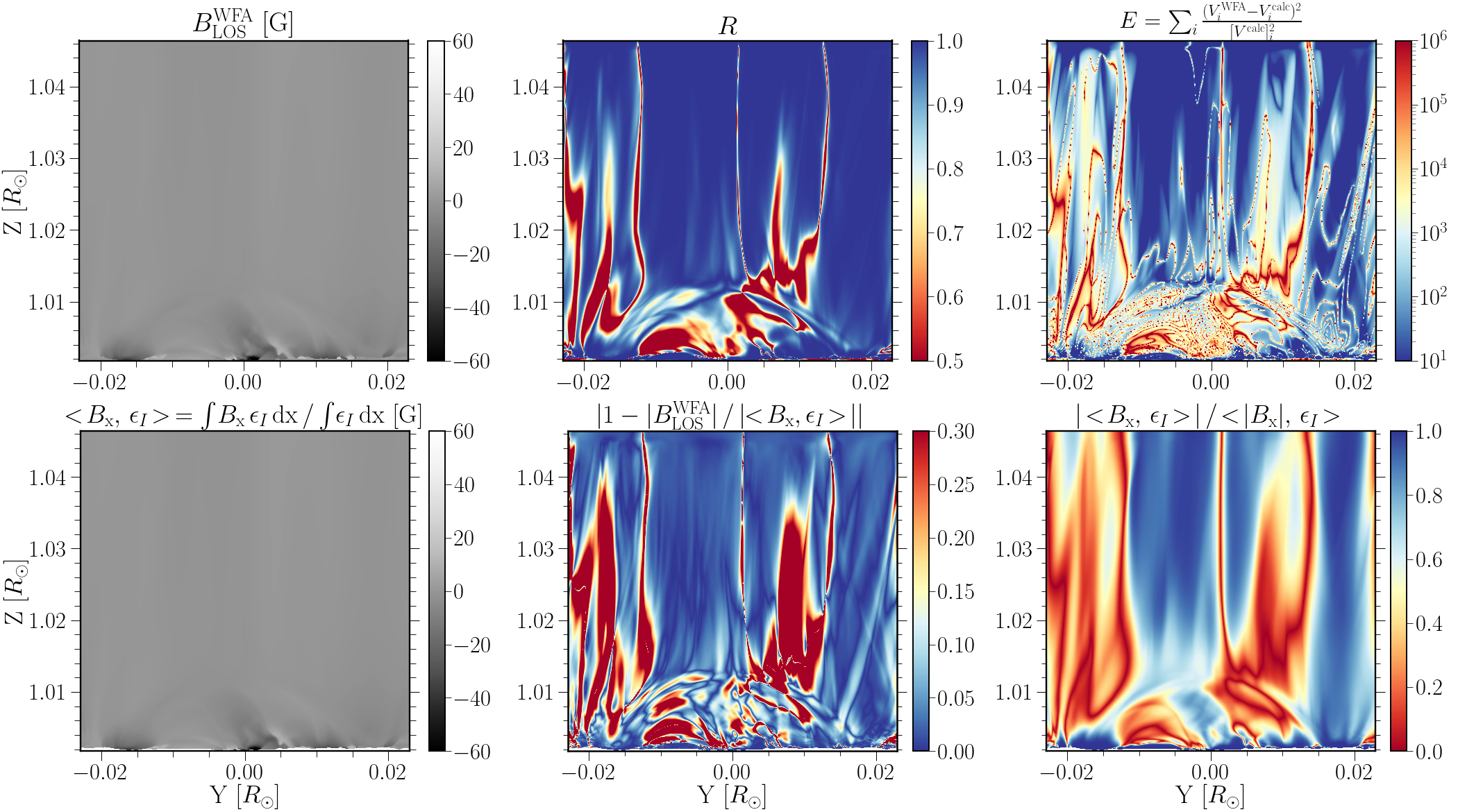}
 \caption{Upper and lower rows: Each panel shows the same quantities as in the corresponding columns of 
 Figs.~\ref{Fig::Multi_WFA} and ~\ref{Fig::Multi_Mod_BLOS}, respectively. 
 Here the quantities are shown for Fe$10747$, 
 but having integrated over the instants of the \gls*{cbp} model between $185$ and $218.33$ minutes (see text).} 
  \label{Fig::Fe13_Stok_WFA_Evol}
\end{figure*}

We also analyzed the information on the magnetic field that can be accessed from the circular polarization signals 
through the \gls*{wfa} (see Section~\ref{sec::WFA}). 
The analogous quantities to those 
corresponding to the Fe$10747$ line in Figures~\ref{Fig::Multi_WFA} and \ref{Fig::Multi_Mod_BLOS}, but for the time-averaged signals, 
are shown in Figure~\ref{Fig::Fe13_Stok_WFA_Evol}. 
The \textcolor{black}{magnetic fields inferred from the \gls*{wfa}}, $B^\mathrm{WFA}_\mathrm{LOS}$ (upper left panel), 
\textcolor{black}{are compared to the emission-weighted fields of the model averaged over the \gls*{los}}, 
$<\!\!B_\mathrm{x}, \epsilon_I\!\!>$ (lower left panel), for which we also averaged over the 
results at different instants in the evolution of the \gls*{cbp} simulation. 
The two values show a reasonably good agreement within the \gls*{cbp}. 
Compared to the results when considering a single instant in time (see Section~\ref{sec::WFA}), 
the time-averaged values of $<\!\!B_\mathrm{x}, \epsilon_I\!\!>$ show a 
noticeable decrease in the average magnitude and a predominance of \textcolor{black}{negative} 
B$_{\mathrm{LOS}}$. 
The decrease in the average magnitude within the \gls*{cbp} is an expected consequence of the change in orientation of the magnetic field within the considered time interval. 
Similarly to the two distinct regions with positive and negative sign found when considering the instant at $201.7$ minutes, 
we attribute the preponderance of a negative sign of $B^\mathrm{WFA}_{\mathrm{LOS}}$ in the \gls*{cbp}  
to the particular orientation of the magnetic field in the region along the \gls*{los} within the strongest emission. 

Regarding the departure from unity of the ratio of $B^\mathrm{WFA}_\mathrm{LOS}$ to $<\!\!B_\mathrm{x}, \epsilon_I\!\!>$  
(lower central panel), we find patches within the \gls*{cbp} region where the 
discrepancies exceed $30\%$ which, similarly to what was discussed in 
Section~\ref{sec::WFA}, tend to coincide with those regions where the inferred magnetic field 
is weakest. As expected, such regions also tend to coincide with those where the Pearson coefficient (upper central panel)
is lowest, the square error $E$ (upper right panel) is largest, and the most cancellations of the 
emission-weighted \gls*{los} occur (lower right panel). 
Thus, although the circular polarization amplitude in the \gls*{cbp} is somewhat attenuated 
when accounting for roughly 30 minutes of evolution in the \gls*{cbp}, it can still provide measurable and useful information about 
the \gls*{los} magnetic fields within the higher-temperature regions of the structure 
from which Fe$10747$ is emitted. 

\section{Conclusions}
\label{sec::Conclusions}

The advent of the new generation of coronagraphs, both in space-based missions such as VELC/Aditya-L1 
\citep[see][]{Singh+19} and large-aperture ground-based facilities such as Cryo-NIRSP/DKIST 
\citep[][]{Rimmele+20,Fehlmann+23}, 
has renewed the interest in studying the intensity and polarization of forbidden spectral lines 
for diagnostics of coronal magnetic fields. 
\textcolor{black}{Here we studied} the suitability of a selection of forbidden lines for 
magnetic diagnostics of fundamental structures in the lower
solar corona, specifically \gls*{cbp}s. We focused on four forbidden lines:
Fe$10747$, Fe$5303$, Si$14301$, and Si$39343$. 

We computed the intensity and polarization patterns 
of the forbidden lines under investigation, considering a model representative of a \gls*{cbp} within a coronal hole, obtained 
from the \textit{Bifrost} simulation presented in \cite{NobregaSiverio+23}. 
The synthetic signals were calculated using the publicly available P-CORONA code \citep{Supriya+25}, 
 which accounts for collisional and radiative processes, scattering polarization, and the 
impact of the magnetic field through the Hanle and Zeeman effect.  
In the calculations, we considered the magnetic fields to be in the Hanle saturation regime. 
For all four considered lines, we used atomic models whose size was 
limited to $31$ levels, 
after verifying the suitability of such models for investigations of \gls*{cbp}s 
using a reduced-resolution atmospheric model for computational feasibility. 

We first focused on a typical instant in the evolution of the \gls*{cbp}, 
corresponding to $201.7$ minutes from the start of the simulation, and computed 
the Stokes profiles. 
As expected, for all four considered lines, the wavelength-integrated intensity 
images showed a clear brightening in the \gls*{cbp} region, 
which extends for some $8$~Mm, or a little in excess of $0.01~ R_\odot$ 
from the base of the corona. The contrast with the fainter intensity emitted from outside the \gls*{cbp} is especially apparent for the 
lines with the highest peak response temperature -- namely Fe$5303$ and Fe$10747$ -- 
and is much more modest for Si$39343$, which has the lowest. 
The circular polarization signals, displayed as $\overline{|V|}/\overline{I\,}$, 
have their largest amplitude within the \gls*{cbp} where, for the four considered lines, 
they approach or exceed $0.1\%$, and 
could be possibly detected using large-aperture coronagraphs with capabilities similar to Cryo-NIRSP/DKIST, but which can observe below 
$0.05\,R_\odot$ from the base of the corona.  

The linear polarization signals, on the other hand, are found to be large outside 
the \gls*{cbp} region (where their amplitude 
exceeds $\sim\!\!1\%$), but they drop 
precipitously within the \gls*{cbp}, generally falling well below $0.1\%$. 
{Underneath the fan surface, where the electron density can reach $10^{9.5}$ cm$^{-3}$, 
collisional processes dominate over radiative ones, reducing the relative contribution of scattering polarization.} 

The radiation emitted from the \gls*{cbp} reaches the observer together with the 
radiation emitted by the enveloping corona along the \gls*{los}. 
To account for this, we added the synthetic profiles resulting from the 
\gls*{cbp} model to those resulting from models representative of the outer 
coronal material, extending from $-3 R_\odot$ to $3 R_\odot$ in the \gls*{los} direction. 
To that end, we used coronal models based on the data provided by Predictive Science, Inc.  
We extracted columns from this model that have the same 
\gls*{pos} dimensions as the \gls*{cbp} model and are tangent to the base of the corona. 
We carried out calculations for two distinct models, PSI1 and PSI2, with the former having 
a slightly lower average temperature and a significantly lower emission in the four considered lines. 
The contribution from the surrounding material, relative to the intensity signal from the \gls*{cbp}, depends both on the peak response 
temperature of the considered line and on the considered PSI model. 
For the Fe$5303$ line, even PSI2 gives only a modest contribution to its signal, due to its high peak response temperature. 
For Fe$10747$, PSI1 does not have a significant impact on the intensity from the \gls*{cbp} which,  
by contrast, is drowned out when considering PSI2. Similar results were found for Si$14301$, although the contribution from the PSI1 model has a 
slightly greater impact on the \gls*{cbp} signals. 
The Si$39343$ signals are completely dominated by the contribution from either model, PSI1 or PSI2.  

{We find that the $\overline{|V|}/\overline{I\,}$ signals are impacted by the surrounding material 
to a similar degree as the intensity.  
This suggests that if the \gls*{cbp} can be clearly observed in intensity, most of the circular polarization should be 
emitted from this structure. On the other {hand}, the contribution of the outer coronal material tends to substantially impact 
the linear polarization amplitude from the \gls*{cbp} for all the considered lines except 
Fe$5303$ (which has a peak response at $2$~MK). 
However, in this case, the linear polarization in the \gls*{cbp} region is quite small, 
being mostly below $0.1\%$ and often on the order of $0.01\%$. 
The suitability of the linear polarization signals for magnetic diagnostics of \gls*{cbp}s 
or similar low-corona structures with 
a large electron density may thus be easily compromised by both their relatively small amplitude and the contribution from external coronal material 
along the \gls*{los}. } 

We also analyzed the application of the \gls*{wfa} to the synthetic intensity and circular polarization 
signals resulting from the \gls*{cbp} model. 
We recall that the \gls*{wfa} mainly provides information about the magnetic field in the 
intervals along the \gls*{los} where the emission is highest for the considered lines. The 
inferred $B^\mathrm{WFA}_\mathrm{LOS}$ is significantly stronger than the average within the \gls*{cbp} box, 
unless we weight the average with the line emissivity along the \gls*{los} (i.e., $<\!B_\mathrm{x}, \epsilon_I\!>$). 
The latter quantity is in good agreement with the inferred $B^{\mathrm{WFA}}_\mathrm{LOS}$  
at the regions on the \gls*{pos}, corresponding to the \gls*{cbp}, where the Pearson 
correlation coefficient between $V(\lambda)$ and $\partial I/\partial\lambda$ is high. 
Such regions coincide with those for which there are no significant cancellations of $B_\mathrm{LOS}$, 
in the intervals along the \gls*{los} where the emission is high. They are more abundant when 
considering lines with a higher peak {response} temperature, most likely because the radiation {is}  
mainly emitted from a narrower spatial region and is thus subject to fewer cancellations. 
Unsurprisingly, the regions with significant cancellations tend to coincide with the regions with 
a small inferred $B_\mathrm{LOS}$. 

The ideal data set to study the \gls*{cbp} magnetic fields should thus present a 
strong intensity contrast between the \gls*{cbp} itself 
and the rest of the coronal material, indicating that the \gls*{cbp} supplies the dominant contribution to the observed signal. 
This would most likely be found in or around a coronal hole. As for the circular polarization, the most 
useful profiles would be those with a high correlation with the 
wavelength derivative of the intensity, typically occurring when there is no magnetic cancellation along the \gls*{los}; 
in those cases one should expect that the \gls*{wfa} can be suitably applied.

Finally, we studied  
how the signals of the Fe$10747$ line change due to the time evolution of the \gls*{cbp}. 
We averaged the signals obtained at different instants of the simulation 
at a $100$s cadence and over an interval of $\sim\!\!30$ minutes  
which, when observing the Fe$10747$ line with large-aperture telescopes, 
would provide a \gls*{snr} large enough to suitably measure polarization signals of about $0.1\%$.  
The resulting time-averaged signals remain suitable for the 
application of the \gls*{wfa}: the \gls*{cbp} is still clearly appreciable in intensity, 
even though the contrast with the surrounding coronal material is somewhat reduced, and there 
is still a reasonably good agreement between the B$_\mathrm{LOS}$ inferred by the \gls*{wfa} and 
the models' emission-weighted B$_\mathrm{LOS}$ averaged along the \gls*{los}. 

The analysis carried out in this paper is particularly relevant for understanding the polarization signals
in systems composed of coronal loops, like the \gls*{cbp}s, which are ubiquitous structures in the solar corona. 
Of the four forbidden lines we investigated, we consider Fe$5303$ and Fe$10747$ to be the most suitable 
for diagnosing CBP magnetic fields. These lines are less affected by contributions from the surrounding coronal 
material and produce the most accurate results when applying the WFA. This is particularly {interesting} given 
that \gls*{cbp}s typically exhibit temperatures between $1.0$ and $3.4$~MK \citep[e.g.,][]{Kariyappa+11, Madjarska19}. 
In such temperature ranges, the hotter lines tend to show higher contrast relative to the ambient corona, especially 
when the CBP is located within a coronal hole or in the quiet Sun -- as is typically the case. Therefore, although our 
conclusions are based on a numerical model, they likely hold in real solar conditions and may be applicable more broadly. 
In addition, our results can provide valuable insights for future missions aiming to perform spectropolarimetric 
observations of structures in the very low corona.  

\vspace{0.3cm}
\begin{center} {\large\textbf{Acknowledgements}} \end{center}
\vspace{-0.2cm}
{This research was supported by the European Research Council through the Synergy grant No. 810218 (``The Whole Sun'' ERC-2018-SyG).
This project also benefited from discussions that took place at the ``Whole Sun Meetings'' supported by the same grant. 
E.A.B. and S.H.D. acknowledge support from the Agencia Estatal de Investigaci\'on 
del Ministerio de Ciencia, Innovaci\'on y Universidades (MCIU/AEI) 
under grant ``Polarimetric Inference of Magnetic Fields'' and the European Regional Development Fund (ERDF) with 
reference PID2022-136563NB-I00/10.13039/501100011033. 
D.N.S. acknowledges the computer resources at the MareNostrum supercomputing installation and 
the technical support provided by the Barcelona Supercomputing Center (BSC, RES-AECT-2021-1-0023, RES-AECT-2022-2-0002), 
as well as the resources provided by Sigma2 - the National Infrastructure for High Performance Computing and Data Storage in Norway. 
Prof. Trujillo Bueno's careful reading of the manuscript and thoughtful comments are recognized with appreciation. 
Illuminating discussions with Drs. de Vicente, del Pino Alem\'an, and Shchukina are gratefully acknowledged. 
Valuable comments from the anonymous referee are sincerely appreciated. }

\bibliographystyle{aa}
\bibliography{cbib}

\begin{appendix}
\onecolumn
\section{Polarization angles}
\label{sec::AppPolang}
\begin{figure*}[ht!]
\centering
\includegraphics[width=0.24\textwidth]{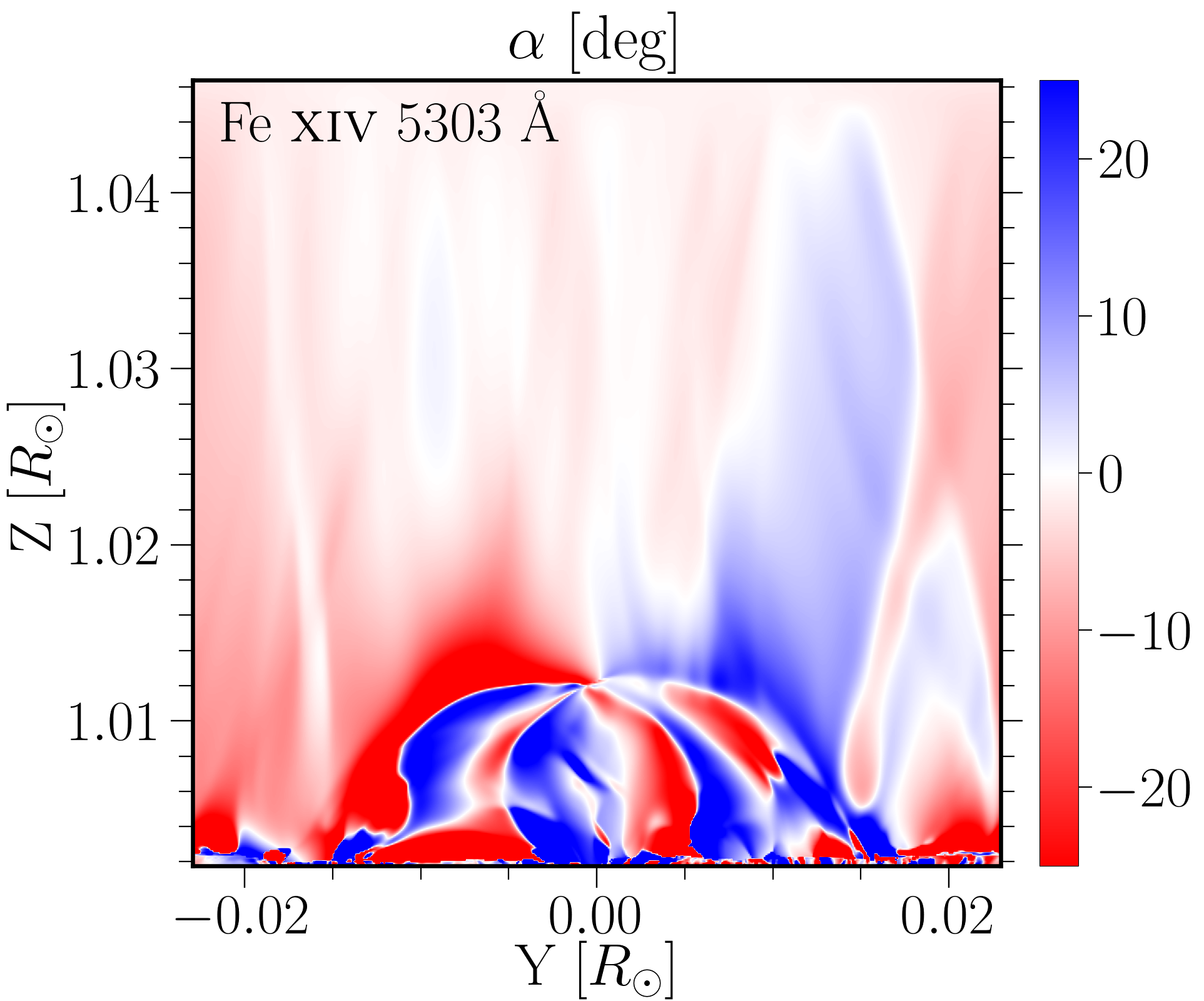}
\includegraphics[width=0.24\textwidth]{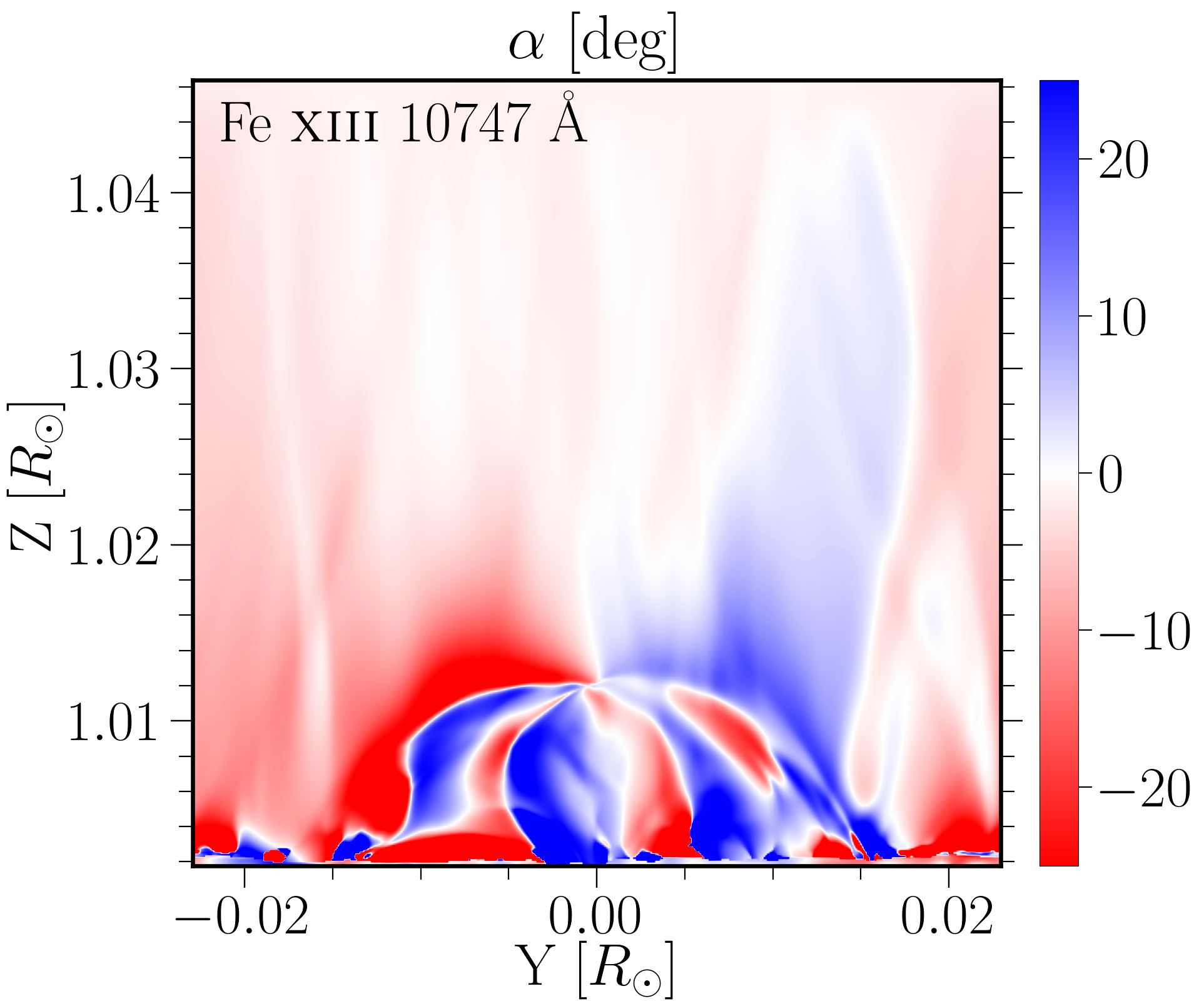}
\includegraphics[width=0.24\textwidth]{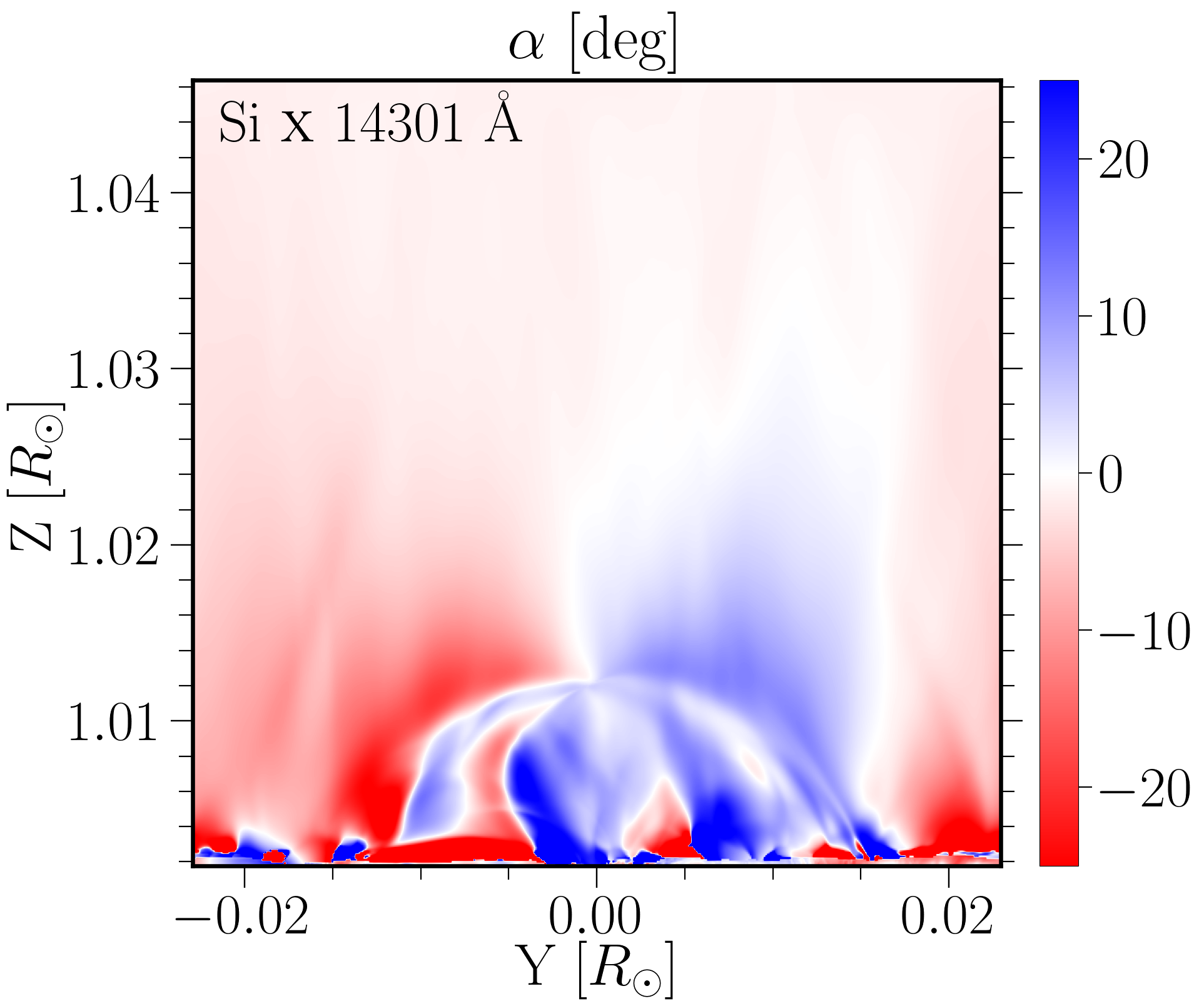}
\includegraphics[width=0.24\textwidth]{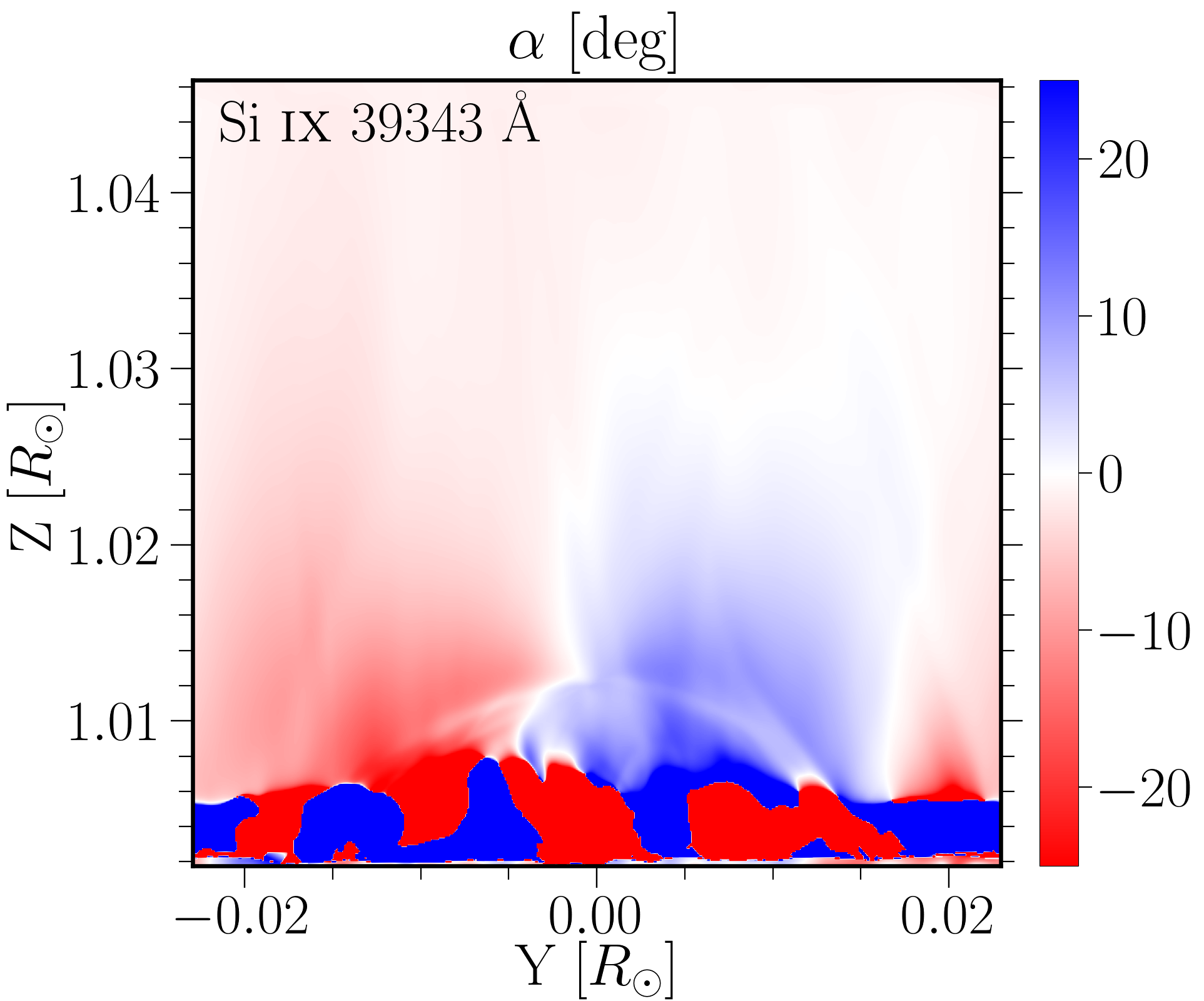}
\caption{Linear polarization angle, with respect to the $Z$ axis, for the radiation emerging from the \gls*{cbp}. 
The four panels correspond to the values for Fe$10747$, Fe$5303$, Si$14301$, and Si$39343$ lines, respectively.  
The intensity and polarization were calculated with P-CORONA, using the \gls*{cbp} model at $201.7$ minutes 
from the start of the simulation. Indeed, such angles were obtained from the same calculations that yielded 
the results shown in Figs.~\ref{Fig::Fe13_First} and \ref{Fig::Multilines_First}, for the corresponding spectral lines. }
\label{Fig::AppPolAngApp}
\end{figure*}

The four panels of Figure~\ref{Fig::AppPolAngApp} show the linear polarization angles $\alpha$ obtained when applying 
P-CORONA for the \gls*{cbp} model, for the Fe$5303$, Fe$10747$, Si$14301$, and Si$39434$ lines, respectively. 
They correspond to the same calculations that produced the $\overline{I\,}$, $\overline{V\,}/\overline{I\,}$ and $P_L$ signals 
presented in Figures~\ref{Fig::Fe13_First} (for Fe$10747$) and \ref{Fig::Multilines_First} (for the other lines). 
The linear polarization angle refers to the angle of the direction of largest linear polarization with respect to the $Z$ axis, with positive (negative) values representing a clockwise (counterclockwise) rotation of the plane of linear polarization (see Equation~\ref{Eq::PolAng}).  

Although the value of $\alpha$ near the edges of the box is slightly affected by the inclination between the
vertical direction, $Z$, and the 
radial direction, this impact is very minor given the small size of the box. 
Indeed, the variation in $\alpha$ is mainly due to the rotation in the plane of polarization induced by 
the magnetic field through the Hanle effect. 
We recall that the Hanle effect operates when the magnetic field is inclined with respect to the symmetry axis of the 
radiation field (in this case, the vertical). Moreover, for \gls*{los}s close to the limb, the direction of Hanle 
rotation depends on the sign of the LoS component of the magnetic field. 
However, in a manner analogous to what was discussed for the circular polarization in Section~\ref{sec::WFA}, the value of $\alpha$ does not simply provide a measure of the average longitudinal magnetic field along the \gls*{los}. The resulting signal gives more weight to %the direction of the magnetic field in the regions from which the emission in linear polarization (rather than intensity) is the strongest. 
regions where the emission in linear polarization (rather than intensity) is the strongest. 

With this in mind, we now compare the $\alpha$ images for the four lines. Outside the \gls*{cbp}, we find the sign to be mostly negative, although there is some spatial fluctuation, especially for Fe$5303$. For the four lines, a plume-like structure with {a} positive sign is produced for $Y > 0$. Within the \gls*{cbp}, there are regions with both positive and negative signs, although with more spatial variability for the Fe$5303$ and Fe$10747$ lines, whose resulting $\alpha$ patterns are very similar. Finally, we highlight the sharp change in the sign for the Si$39343$ close to the base of the corona, which we attribute to the change in the sign of atomic polarization for electron densities around $10^{8.5}$ cm$^{-3}$ reported by \cite{SchadDima20}. 

Unfortunately, it must be stressed that the suitability for magnetic diagnostics of the linear polarization (including its angle) from
the \gls*{cbp} is compromised by its relatively small amplitude and the contribution from the surrounding material. 
These drawbacks may be less of {a concern} for the linear polarization signals outside the \gls*{cbp}, 
but a more detailed analysis is beyond the scope of the present work.

\section{Time evolution of the Fe$10747$ signals in the CBP}

\begin{figure*}[ht!]
\centering

\caption{Animation available in the online material, which displays the Fe$10747$ signals calculated with P-CORONA, as detailed in Section~\ref{sec::Formulation}. The calculations were carried out using the CBP model introduced in Section~\ref{sec::FormulSubAtmosSSubCBP}, but at different instants between 183 and 218.3 minutes, in increments of $\sim$1.6 minutes (i.e., 100 seconds). \textit{Left panel}: Wavelength-integrated intensity $\overline{I\,}$. \textit{Central panel}: Wavelength-integrated absolute-value circular polarization, normalized to the integrated intensity, $\overline{|V|}/\overline{I\,}$. \textit{Right panel}: Wavelength-integrated linear polarization fraction $P_L$.}
\end{figure*}

% \label{LastPage} % To remove superfluous warnings, see also edits to aa.cls
{\textcolor{white}{\pageref*{LastPage}}}

\end{appendix}
\end{document}